\documentclass[12pt, final, onecolumn]{IEEEtran}
\usepackage{graphicx}  
\usepackage{newtxmath}      
\usepackage{amsmath}
\usepackage{bm}
\usepackage{cite}

\newcommand{\nv}{{\bf n}}

\newcommand{\pv}{{\bf p}}

\newcommand{\sv}{{\bf s}}

\newcommand{\xv}{{\bf x}}
\newcommand{\yv}{{\bf y}}

\newcommand{\onev}{{\bf 1}}

\newcommand{\thetav}{{\boldsymbol \theta}}

\newcommand{\Id}{{\bf I}}

\newcommand{\Ac}{{\cal A}}

\newcommand{\Cc}{{\cal C}}

\newcommand{\Mc}{{\cal M}}
\newcommand{\Nc}{{\cal N}}

\newcommand{\EE}{\mathbb{E}}

\newcommand{\RR}{\mathbb{R}}

\newcommand{\LB}{\left(}
\newcommand{\RB}{\right)}
\newcommand{\LSB}{\left[}
\newcommand{\RSB}{\right]}

\begin{document}

\title{Deep Learning for Wireless Communications}

\author{Tugba~Erpek, Timothy~J.~O'Shea, Yalin~E.~Sagduyu, Yi~Shi,  T.~Charles~Clancy

\thanks{\newline T. Erpek and T. C. Clancy are with Virginia Tech, Arlington, VA, USA. E-mail: \{terpek,tcc\}@vt.edu \newline
T. J. O'Shea is with Virginia Tech, Arlington, VA and DeepSig, Inc., Arlington, VA, USA. E-mail: oshea@vt.edu  \newline
Y. E. Sagduyu is with Intelligent Automation, Inc., Rockville, MD, USA. E-mail: ysagduyu@i-a-i.com  \newline
Y. Shi is with Virginia Tech, Blacksburg, VA, USA. 
E-mail: yshi@vt.edu}}


\maketitle

\begin{abstract}
Existing communication systems exhibit inherent limitations in translating theory to practice when handling the complexity of optimization for emerging wireless applications with high degrees of freedom. Deep learning has a strong potential to overcome this challenge via data-driven solutions and improve the performance of wireless systems in utilizing limited spectrum resources. In this chapter, we first describe how deep learning is used to design an end-to-end communication system using autoencoders. This flexible design effectively captures channel impairments and optimizes transmitter and receiver operations jointly in single-antenna, multiple-antenna, and multiuser communications. Next, we present the benefits of deep learning in spectrum situation awareness ranging from channel modeling and estimation to signal detection and classification tasks. Deep learning improves the performance when the model-based methods fail. Finally, we discuss how deep learning applies to wireless communication security. In this context, adversarial machine learning provides novel means to launch and defend against wireless attacks. These applications demonstrate the power of deep learning in providing novel means to design, optimize, adapt, and secure wireless communications.  
\end{abstract}


\IEEEpeerreviewmaketitle

\section{Introduction} \label{sec:intro}
It is of paramount importance to deliver information in wireless medium from one point to another quickly, reliably, and securely. Wireless communications is a field of rich expert knowledge that involves designing waveforms (e.g., long-term evolution (LTE) and fifth generation mobile communications systems (5G)), modeling channels (e.g., multipath fading), handling interference (e.g., jamming) and traffic (e.g., network congestion) effects, compensating for radio hardware imperfections (e.g., RF front end non-linearity), developing communication chains (i.e., transmitter and receiver), recovering distorted symbols and bits (e.g., forward error correction), and supporting wireless security (e.g., jammer detection). The design and implementation of conventional communication systems are built upon strong probabilistic analytic models and assumptions. However, existing communication theories exhibit strong limitations in utilizing limited spectrum resources and handling the complexity of optimization for emerging wireless applications (such as spectrum sharing, multimedia, Internet of Things (IoT), virtual and augmented reality), each with high degrees of freedom. Instead of following a rigid design, new generations of wireless systems empowered by cognitive radio \cite{Haykin2005} can learn from spectrum data, and optimize their spectrum utilization to enhance their performance. These smart communication systems rely on various detection, classification, and prediction tasks such as signal detection and signal type identification in spectrum sensing to increase situational awareness. To achieve the tasks set forth in this vision, machine learning (especially deep learning) provides powerful automated means for communication systems to learn from spectrum data and adapt to spectrum dynamics \cite{Clancy2007}. 

\begin{figure}[h]
\centering
\includegraphics[width=0.99\textwidth,trim={0cm 0cm 0cm 0cm},clip]{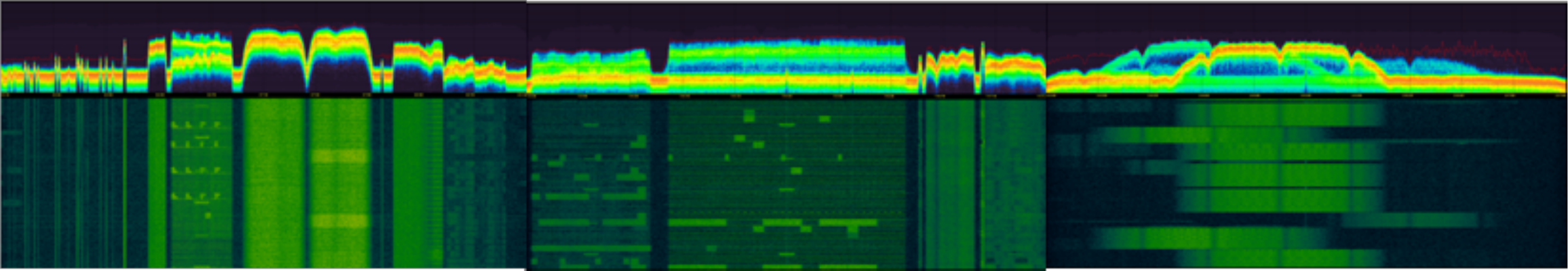}
\caption{Example of dynamic spectrum data from several frequency bands.}
\label{fig:spectrumdata}
\end{figure}

\begin{figure}[h]
\centering
\includegraphics[width=0.5\textwidth]{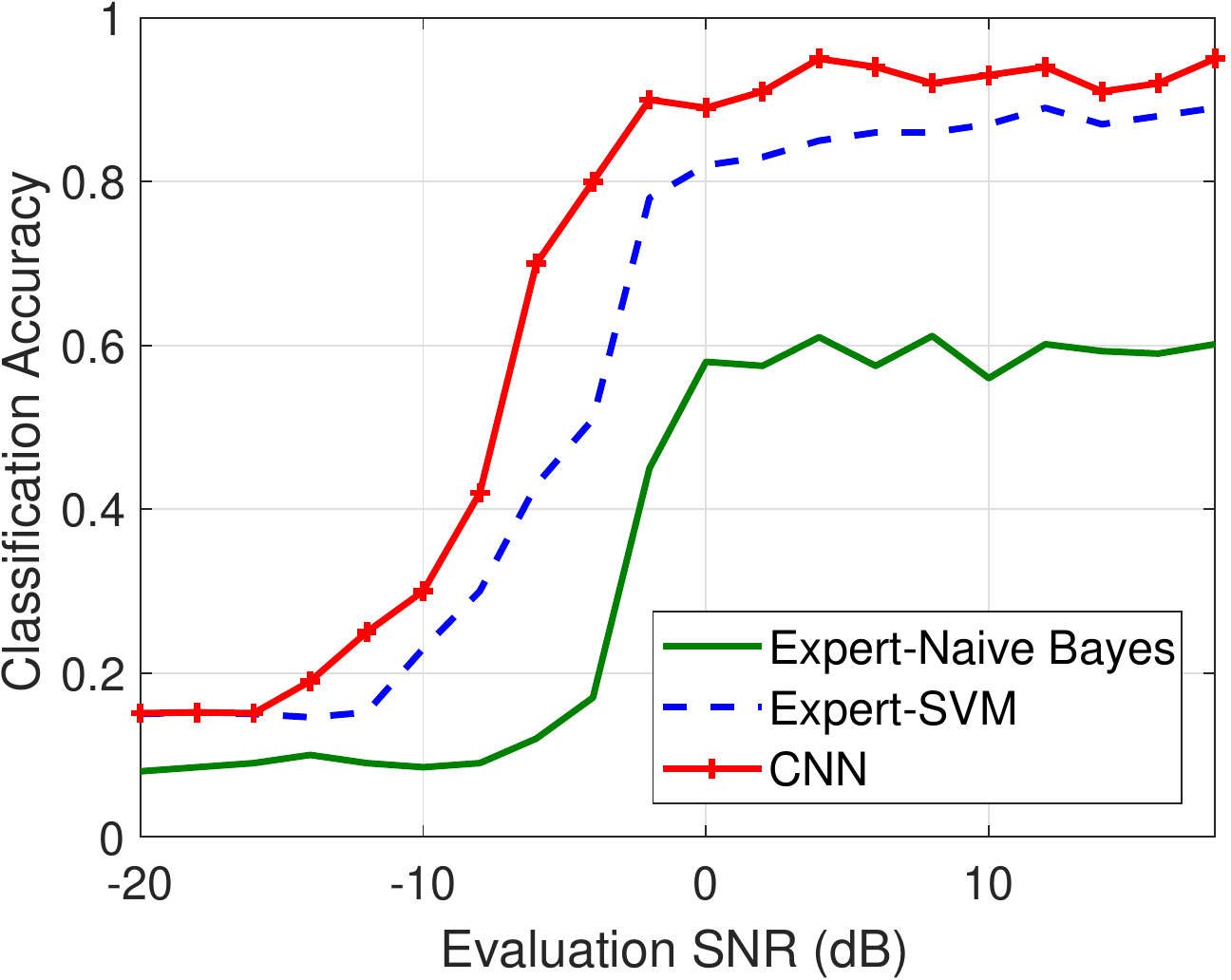}
\caption{Example of deep learning outperforming conventional machine learning in wireless domain. CNN is more successful than SVM and Naive Bayes in classifying a variety of digital modulations (BPSK, QPSK, 8PSK, 16-QAM, 64-QAM, BFSK, CPFSK, and PAM4) and analog modulations (WB-FM, AM-SSB, and AM-DSB) \cite{RecNetworks}.}
\label{fig:DLvsML}
\end{figure}

Wireless communications combine various waveform, channel, traffic, and interference effects, each with its own complex structures that quickly change over time, as illustrated in Fig.~\ref{fig:spectrumdata}. The data underlying wireless communications come in large volumes and at high rates, e.g., gigabits per second in 5G, and is subject to harsh interference and various security threats due to the shared nature of wireless medium. Traditional modeling and machine learning techniques often fall short of capturing the delicate relationship between highly complex spectrum data and communication design, while deep learning has emerged as a viable means to meet data rate, speed, reliability, and security requirements of wireless communication systems. One motivating example in this regard is from signal classification where a receiver needs to classify the received signals \cite{RecNetworks} based on waveform features, e.g., modulation used at the transmitter that adds the information to the carrier signal by varying its properties (e.g., amplitude, frequency, or phase). This signal classification task is essential in dynamic spectrum access (DSA) where a transmitter (secondary user) needs to first identify signals of primary users (such as TV broadcast networks) who has the license to operate on that frequency and then avoid interference with them (by not transmitting at the same time on the same frequency). 
Fig.~\ref{fig:DLvsML} shows that deep learning based on convolutional neural networks (CNN) achieves significantly higher accuracy in signal classification compared to feature based classifiers using support vector machine (SVM) or Naive Bayes. This performance gain is consistent across different signal-to-noise ratio (SNR) levels that capture the distance from transmitter to receiver and the transmit power. One particular reason is that conventional machine learning algorithms rely on the representative value of inherent features that cannot be reliably extracted from spectrum data, where deep learning can be readily applied to raw signals and can effectively operate using feature learning and latent representations.

This chapter presents methodologies and algorithms to apply deep learning to wireless communications in three main areas.

\begin{enumerate}
\item Deep learning to design \emph{end-to-end (physical layer) communication chain} (Sec. \ref{sec:endtoend}).
\item Deep learning to support \emph{spectrum situation awareness} (Sec. \ref{sec:sitawar}).
\item Deep learning for \emph{wireless security} to launch and defend wireless attacks (Sec. \ref{sec:security}). 
\end{enumerate}

In Sec. \ref{sec:endtoend}, we formulate an end-to-end physical layer communications chain (transmitter and receiver) as an \emph{autoencoder} that is based on two \emph{deep neural networks} (DNNs), namely an \emph{encoder} for the transmitter functionalities such as modulation and coding, and a \emph{decoder} for the receiver functionalities such as demodulation and decoding. By incorporating the channel impairments in the design process of autoencoder, we demonstrate the performance gains over conventional communication schemes.  In Sec. \ref{sec:sitawar}, we present how to use different DNNs such as \emph{feedforward}, \emph{convolutional}, and \emph{recurrent neural networks} for a variety of spectrum awareness applications ranging from channel modeling and estimation to spectrum sensing and signal classification. To support fast response to spectrum changes, we discuss the use of \emph{autoencoder} to extract latent features from wireless communications data and the use of \emph{generative adversarial networks} (GANs) for spectrum data augmentation to shorten spectrum sensing period. Due to the open and broadcast nature of wireless medium, wireless communications are prone to various attacks such as jamming. In Sec. \ref{sec:security}, we present emerging techniques built upon \emph{adversarial deep learning} to gain new insights on how to attack wireless communication systems more intelligently compared to conventional wireless attack such as jamming data transmissions. We also discuss a defense mechanism where the adversary can be fooled when adversarial deep learning is applied by the wireless system itself.

\section{Deep Learning for End-to-end Communication Chain} \label{sec:endtoend}
\begin{figure}[h]
\centering
\includegraphics[width=0.9\textwidth,trim={0cm 1cm 0cm 1cm},clip]{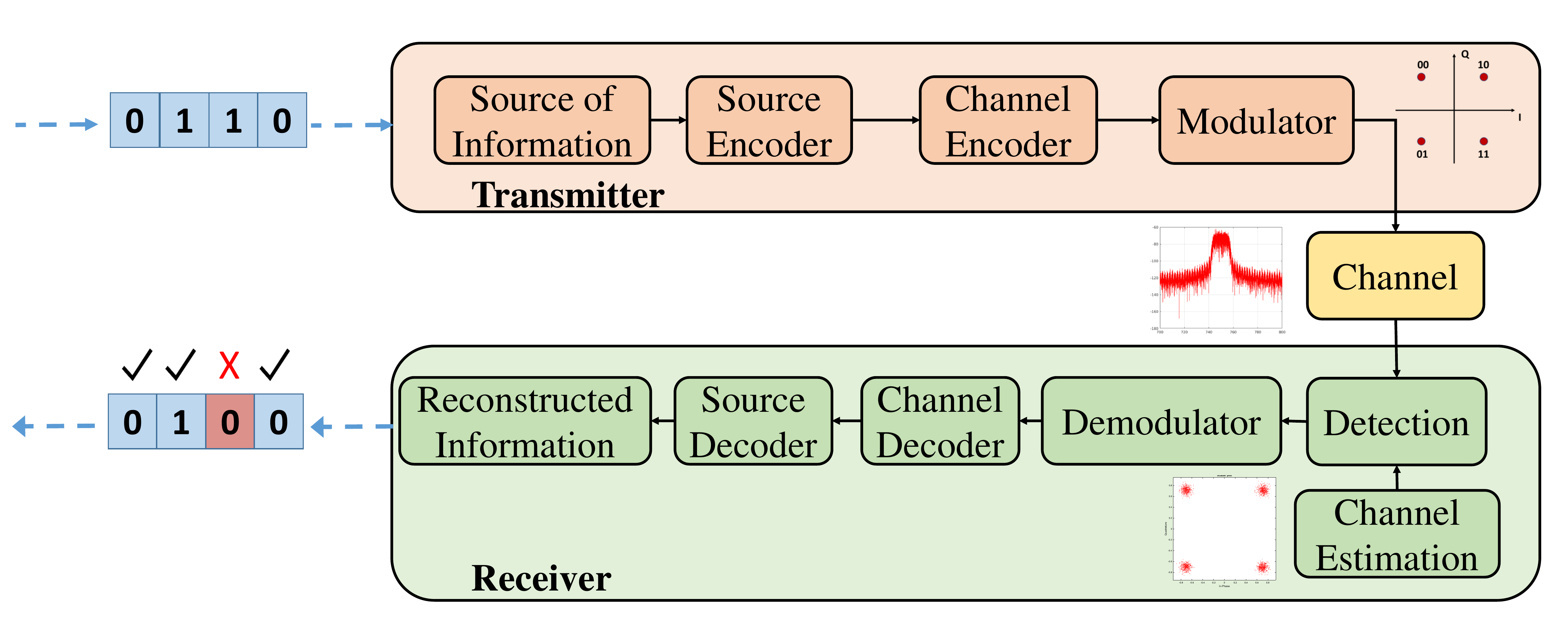}
\caption{Conventional communication system block diagram.}
\label{fig:CommSys}
\end{figure}

The fundamental problem of communication systems is to transmit a message such as a bit stream from a transmitter using radio waves and reproduce it either exactly or approximately at a receiver \cite{Shannon}. The focus in this section is on the physical layer of the Open Systems Interconnection (OSI) model. Conventional communication systems split signal processing into a chain of multiple independent blocks separately at the transmitter and receiver, and optimize each block individually for a different functionality. Fig.~\ref{fig:CommSys} shows the block diagram of a conventional communication system. The source encoder compresses the input data and removes redundancy. Channel encoder adds redundancy on the output of the source encoder in a controlled way to cope with the negative effects of the communication medium. Modulator block changes the signal characteristics based on the desired data rate and received signal level at the receiver (if adaptive modulation is used at the transmitter). The communication channel distorts and attenuates the transmitted signal. Furthermore, noise is added to the signal at the receiver due to the receiver hardware impairments. Each communication block at the transmitter prepares the signal to the negative effects of the communication medium and receiver noise while still trying to maximize the system efficiency. These operations are reversed at the receiver in the same order to reconstruct the information sent by the transmitter. This approach has led to efficient, versatile, and controllable communication systems that we have today with individually optimized processing blocks. However, this individual optimization process does not necessarily optimize the overall communication system. For example, the separation of source and channel coding (at the physical layer) is known to be sub-optimal \cite{GoldsmithVTC}. The benefit in joint design of communication blocks is not limited to physical layer but spans other layers such as medium access control at link layer and routing at network layer \cite{SagduyuIT2008}. Motivated by this flexible design paradigm, deep learning provides automated means to treat multiple communications blocks at the transmitter and the receiver jointly by training them as combinations of DNNs.  

MIMO systems improve spectral efficiency by using multiple antennas at both transmitter and receiver to increase the communication range and data rate. Different signals are transmitted from each antenna at the same frequency. Then each antenna at the receiver receives superposition (namely, interference) of the signals from transmitter antennas in addition to the channel impairments (also observed for single antenna systems). The traditional algorithms developed for MIMO signal detection are iterative reconstruction approaches and their computational complexity is impractical for many fast-paced applications that require effective and fast signal processing to provide high data rates \cite{SamuelDeepMIMO,WangSurvey}. Model-driven MIMO detection techniques can be applied to optimize the trainable parameters with deep learning and improve the detection performance. As an example, a MIMO detector was built in \cite{SamuelDeepMIMO} by unfolding a projected gradient descent method. The deep learning architecture used a compressed sufficient statistic as an input in this scheme. Another model-driven deep learning network was used in \cite{HeMIMODet} for the orthogonal approximate message passing algorithm. 

Multiuser communication systems, where multiple transmitters and/or receivers communicate at the same time on the same frequency, allow efficient use of the spectrum, e.g., in an \textit{interference channel} (IC), multiple transmitters communicate with their intended receivers on the same channel. The signals received from unintended transmitters introduce additional interference which needs to be eliminated with precoding  at the transmitters and signal processing at the receivers. The capacity region for IC in \textit{weak}, \textit{strong} and \textit{very strong interference regimes} has been studied extensively \cite{Veeravalli, HanKobayashi, Carleial}. Non-orthogonal multiple access (NOMA) has emerged to improve the spectral efficiency by allowing some degree of interference at receivers that can be efficiently controlled across interference regimes \cite{ErpekICNC}. However, the computational complexity of such capacity-achieving schemes is typically high to be realized in practical systems.    

Recently, deep learning-based end-to-end communication systems have been developed for single antenna \cite{OSheaTCCN,DornerOTA}, multiple antenna \cite{OSheaAllerton}, and multiuser \cite{OSheaTCCN, ErpekICC} systems to improve the performance of the traditional approaches by jointly optimizing the transmitter and the receiver as an \textit{autoencoder}  instead of optimizing individual modules both at the transmitter and receiver. Autoencoder is a DNN that consists of an encoder that learns a (latent) representation of the given data and a decoder that reconstructs the input data from the encoded data  \cite{Goodfellow_2016_Book}. In this setting, joint modulation and coding at the transmitter corresponds to the encoder, and joint decoding and demodulation at the receiver corresponds to the decoder. The joint optimization includes multiple transmitter and receivers for the multiuser case to learn and eliminate the additional interference caused by multiple transmitters. The following sections will present the autoencoder-based communication system implementations and their performance evaluation.

\subsection{Single Antenna Systems}\label{subsec:singleant}
\begin{figure}[h]
	\centering
	\begin{tabular}{c}
		\includegraphics[width=0.3\textwidth, trim={4cm 4cm 4cm 3cm},clip]{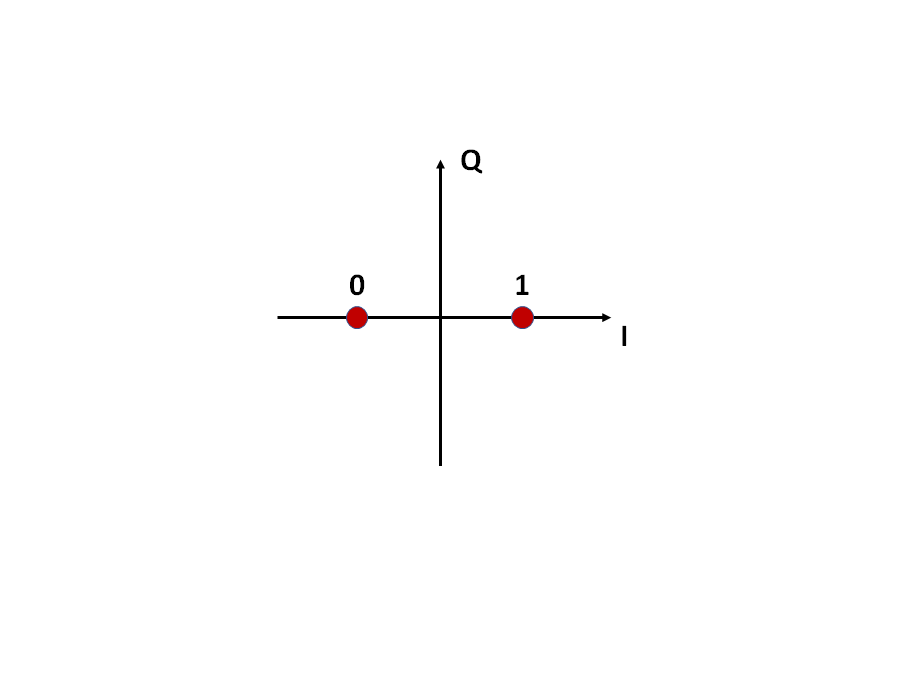}
		\includegraphics[width=0.3\textwidth, trim={4cm 4cm 4cm 3cm},clip]{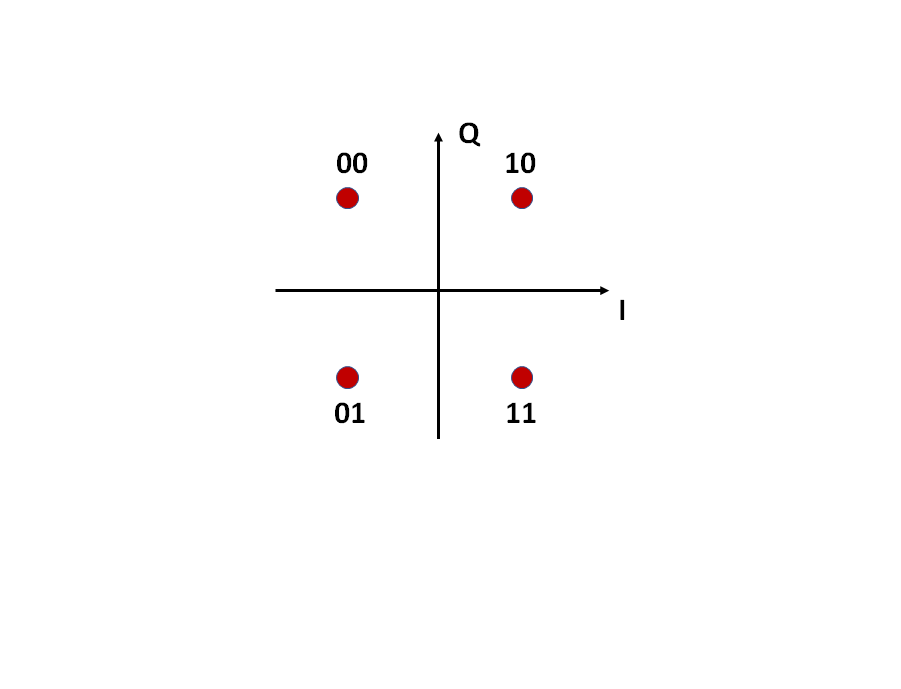}
		\includegraphics[width=0.3\textwidth, trim={0.5cm 0.5cm 0.5cm 0.5cm},clip]{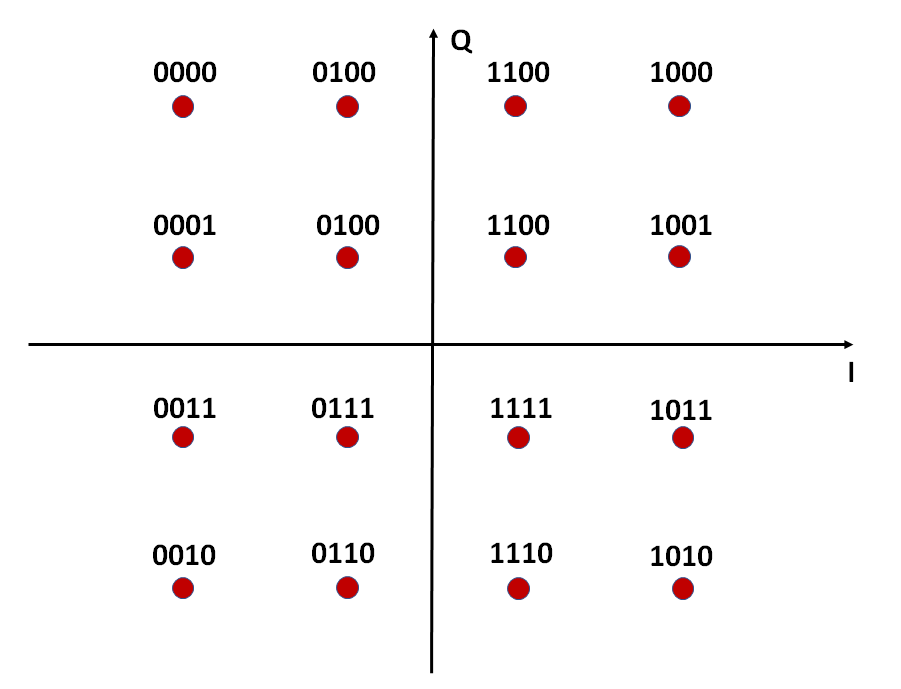}
		\\[-5pt]
		\footnotesize{\hskip0mm(a)} \hskip47.8mm \footnotesize{(b)} \hskip47.8mm \footnotesize{(c)}
	\end{tabular}
	\caption{Example of digital modulation constellations (a) BPSK, (b) QPSK, (c) 16-QAM.}
	\label{fig:constBPSKQPSKQAM}
\end{figure}

A communication system consists of a transmitter, a receiver, and channel that carries information from the transmitter to the receiver. A fundamental new way to think about communication system design is to formulate it as an end-to-end reconstruction task that seeks to jointly optimize transmitter and receiver components in a single process using autoencoders \cite{OSheaTCCN}. As in the conventional communication systems, the transmitter wants to communicate one out of $M$ possible messages $s\in \Mc=\{1,2,...,M\}$ to the receiver making $n$ discrete uses of the channel. It applies the modulation process $f: \Mc \mapsto \RR^{n}$ to the message $s$ to generate the transmitted signal $\xv=f(s)\in\RR^{n}$. The input symbols from a discrete alphabet are mapped to the points (complex numbers) on the constellation diagram as part of digital modulation. The digital modulation schemes for conventional communication systems have pre-defined constellation diagrams. The symbols are constructed by grouping the input bits based on the desired data rate. The desired data rate determines the constellation scheme to be used. Fig.~\ref{fig:constBPSKQPSKQAM} shows the constellation diagrams for the binary phase shift keying (BPSK), quadrature phase shift keying (QPSK), and 16-quadrature amplitude modulation (QAM) as example of digital modulation schemes and their symbol mapping. Linear decision regions make the decoding task relatively simpler at the receiver. For the autoencoder system, the output constellation diagrams are not pre-defined. They are optimized based on the desired performance metric, i.e.,  the symbol error rate to be reduced at the receiver.    

The hardware of the transmitter imposes an energy constraint $\lVert \xv \rVert_2^2 \le n$, amplitude constraint $| x_i | \le 1\, \forall i$, or an average power constraint $\EE\LSB |x_i|^2 \RSB\le 1\,\forall i $ on $\xv$. The data rate of this communication system is calculated as $R=k/n$\,[bit/channel~use], where $k=\log_2(M)$ is the number of input bits and $n$ can be considered as the output of a forward error correction scheme where it includes both the input bits and redundant bits to mitigate the channel effects. As a result, the notation ($n$,$k$) means that a communication system sends one out of $M=2^k$ messages (i.e., $k$ bits) through $n$ channel uses. The communication channel is described by the conditional probability density function $p(\yv|\xv)$, where $\yv\in\RR^n$ denotes the received signal. Upon reception of $\yv$, the receiver applies the transformation $g:\RR^n \mapsto \Mc$ to produce the estimate $\hat{s}$ of the transmitted message $s$.
Mapping $\xv$ to $\yv$ is optimized in a \emph{channel autoencoder} so that the transmitted message can be recovered with a small probability of error. In other words, autoencoders used in many other deep learning application areas typically remove redundancy from input data by compressing it; however, the channel autoencoder adds controlled redundancy to learn an intermediate representation robust to channel perturbations.

\begin{figure*}
\centering
\includegraphics[width=0.7\textwidth, trim={4cm 3cm 4cm 4.5cm},clip]{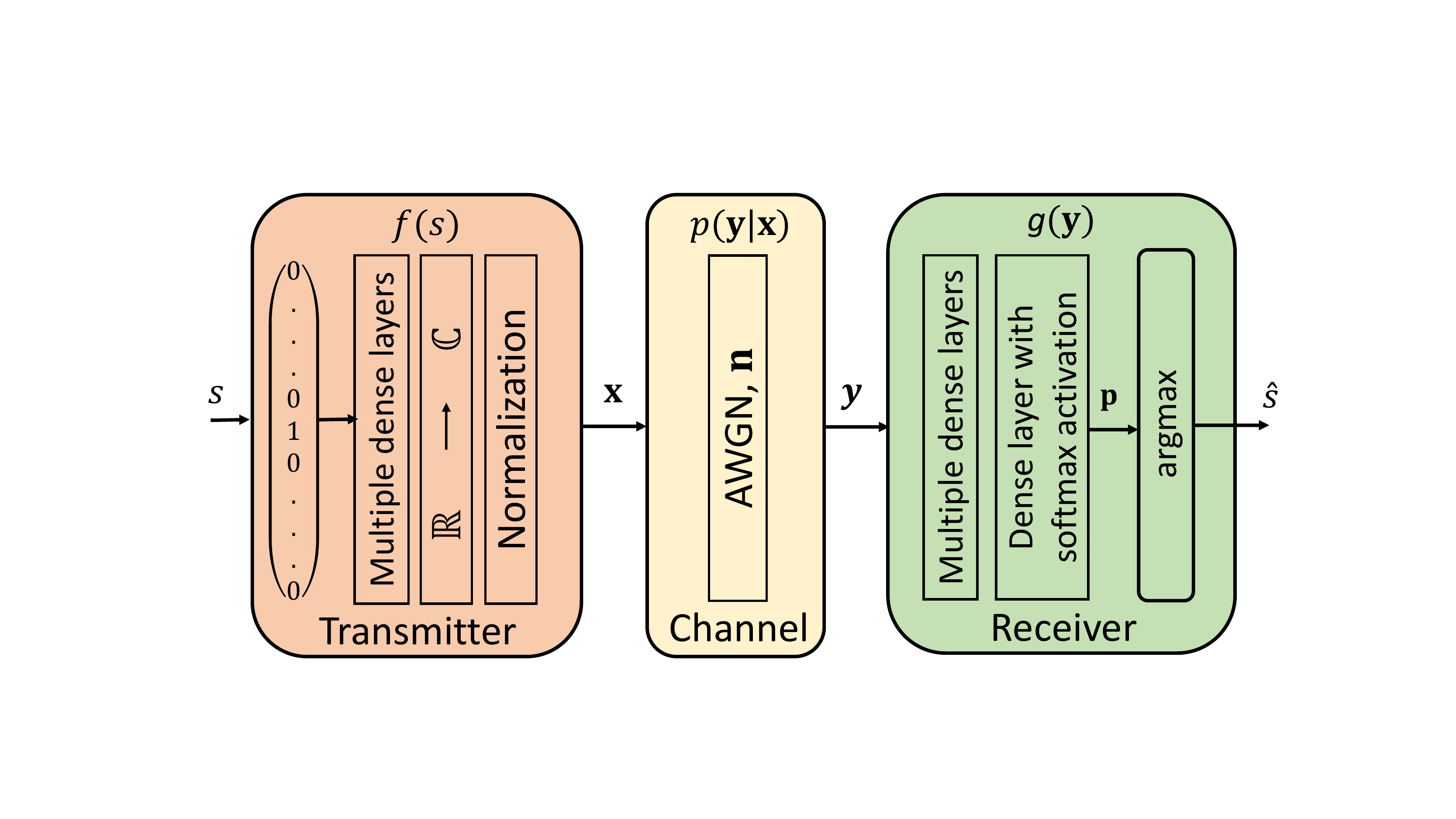}
\caption{A communication system over an additive white Gaussian noise (AWGN) channel represented as an autoencoder. The input $s$ is encoded as a one-hot vector, the output is a probability distribution over all possible messages. The message with the highest probability is selected as output $\hat{s}$. \label{fig:ae_awgn}}
\end{figure*}

The block diagram of the channel autoencoder scheme is shown in Fig.~\ref{fig:ae_awgn}. The input symbol is represented as a one-hot vector. The transmitter consists of a feedforward neural network (FNN) with multiple dense layers. The output of the last dense layer is reshaped to have two values that represent complex numbers with real (in-phase, I) and imaginary (quadrature, Q) parts for each modulated input symbol. The normalization layer ensures that physical constraints on $\xv$ are met. The channel is represented by an additive noise layer with a fixed variance $\beta=(2RE_b/N_0)^{-1}$, where $E_b/N_0$ denotes the energy per bit ($E_b$) to noise power spectral density ($N_0$) ratio. The receiver is also implemented as an FNN. Its last layer uses a softmax activation whose output $\pv\in(0,1)^{M}$ is a probability vector over all possible messages. The index of the element of $\pv$ with the highest probability is selected as the decoded message. The autoencoder is trained using stochastic gradient descent (SGD) algorithm on the set of all possible messages $s\in\Mc$ using the well suited categorical cross-entropy loss function between $\onev_s$ and $\pv$. The noise value changes in every training instance. Noise layer is used in the forward pass to distort the transmitted signal. It is ignored in the backward pass.  

\begin{figure}[h]
    \centering
    \begin{tabular}{c}
    \includegraphics[width=0.45\textwidth]{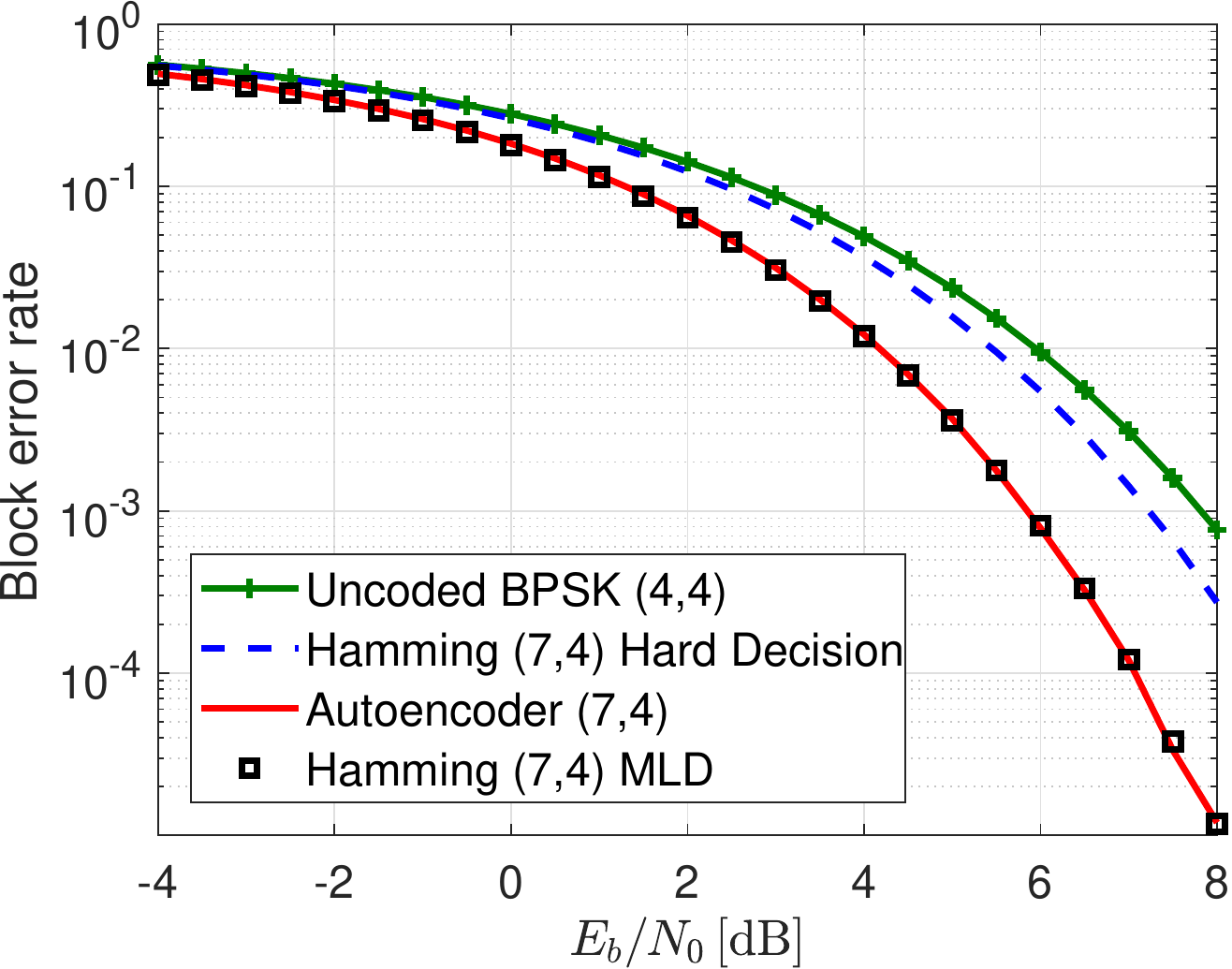} \hspace*{2em}
    \includegraphics[width=0.45\textwidth]{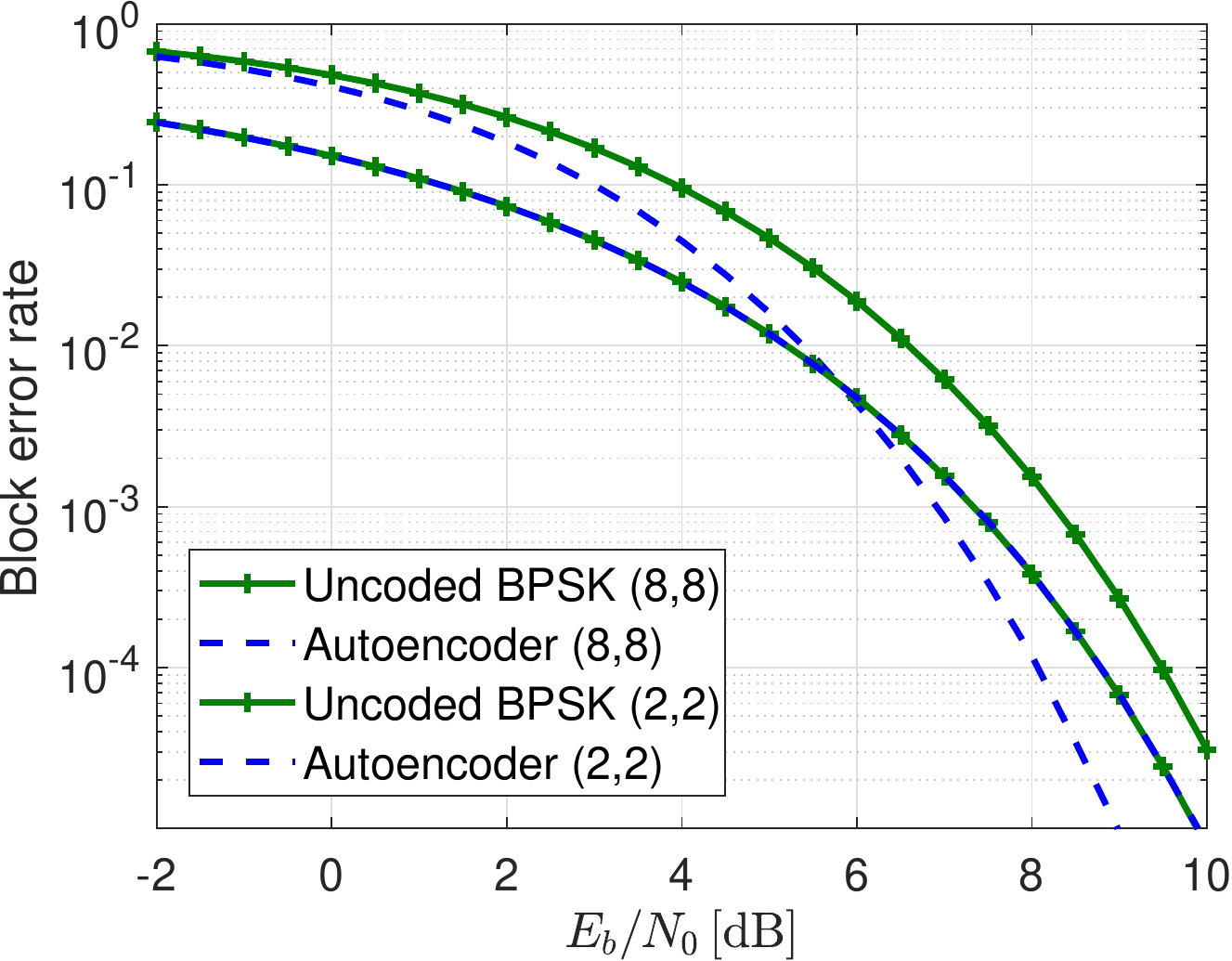}
    \\[-5pt]
    \footnotesize{\hskip9mm(a)} \hskip86.8mm \footnotesize{(b)} 
    \end{tabular}
\caption{BLER versus $E_b/N_0$ for the autoencoder and several baseline communication schemes \cite{OSheaTCCN}.}
\label{fig:nnvsHamming_BPSK}
\end{figure}

Fig.~\ref{fig:nnvsHamming_BPSK} (a) compares the block error rate (\Ac{BLER}), i.e., ${\Pr(\hat{s}\neq s)}$, of a communication system employing BPSK modulation and a Hamming~(7,4) code with either binary hard-decision decoding or maximum likelihood  decoding (MLD) against the \Ac{BLER} achieved by the trained autoencoder~(7,4) (with fixed energy constraint $\lVert \xv \rVert_2^2 = n$). Autoencoder is trained at $E_b/N_0=7\,$dB using Adam \cite{AdamOpt} optimizer with learning rate $0.001$. Both systems operate at rate $R=4/7$. The \Ac{BLER} of uncoded BPSK~(4,4) is also included for comparison. The autoencoder learns encoder and decoder functions without any prior knowledge that achieve the same performance as the Hamming~(7,4) code with MLD. Table~\ref{tab:ae-layout} shows the number of neural network layers used at the encoder (transmitter) and decoder (receiver) of the autoencoder system. 

\begin{figure}[h]
	\centering
	\begin{tabular}{c}
		\includegraphics[width=0.25\columnwidth]{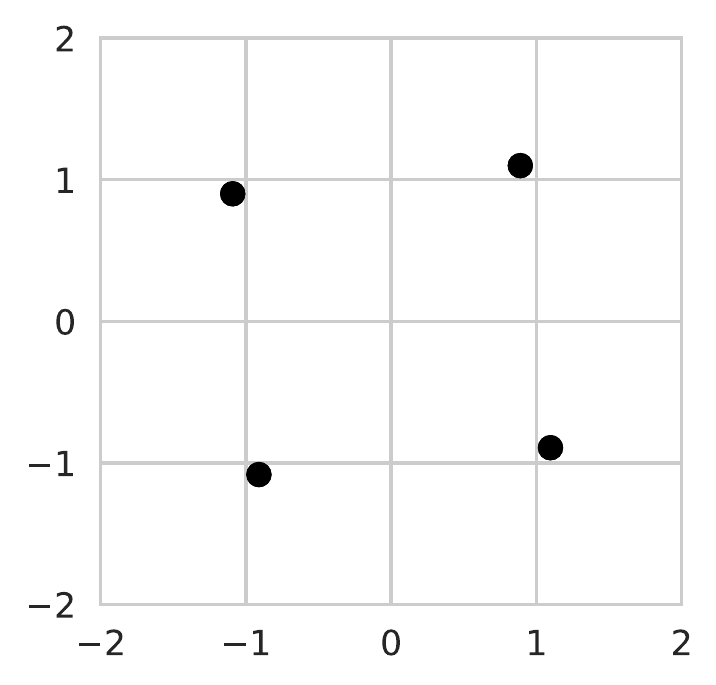} \hspace*{1em}
		\includegraphics[width=0.25\columnwidth]{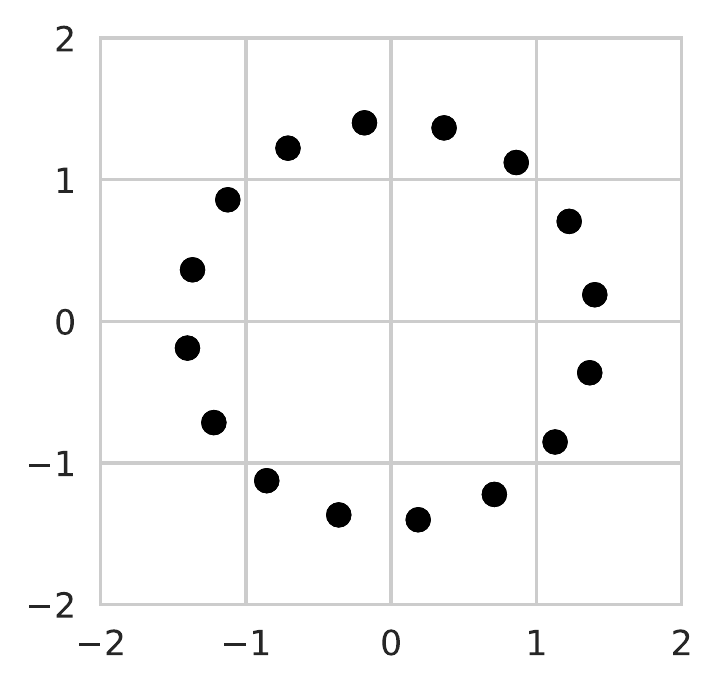} \hspace*{1em}
		\includegraphics[width=0.25\columnwidth]{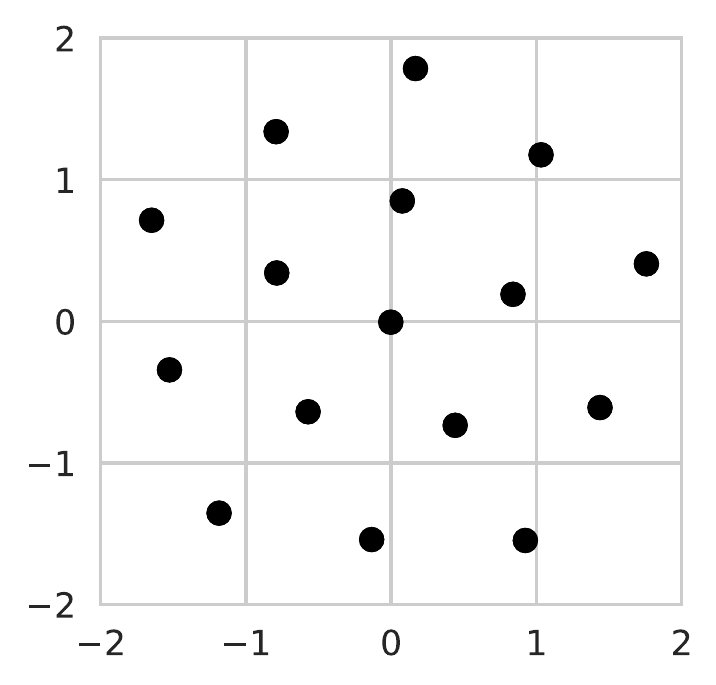}
		\\[-5pt]
		\footnotesize{\hskip5mm(a)} \hskip47.8mm \footnotesize{(b)} \hskip47.8mm \footnotesize{(c)}
	\end{tabular}
	\caption{Constellations produced by autoencoders using parameters $(n,k)$: (a) $(2,2)$ (b) $(2,4)$, (c) $(2,4)$ with average power constraint.}\label{fig:ae_consts}
\end{figure}

Fig.~\ref{fig:nnvsHamming_BPSK} (b) shows the performance curves for (8,8) and (2,2) communication systems when $R=1$. The autoencoder achieves the same \Ac{BLER} as uncoded BPSK for (2,2) system and it outperforms the latter for (8,8) system, implying that it has learned a joint coding and modulation scheme, such that a coding gain is achieved. Fig.~\ref{fig:ae_consts} shows the constellations $\xv$ of all messages for different values of $(n,k)$ as complex constellation points, i.e., the $x$- and $y$-axes correspond to the first and second transmitted symbols, respectively. Fig.~\ref{fig:ae_consts} (a) shows the simple $(2,2)$ system that converges rapidly to a classical \Ac{QPSK} constellation (see Fig.~\ref{fig:constBPSKQPSKQAM} (b)) with some arbitrary rotation.  Similarly, Fig.~\ref{fig:ae_consts} (b) shows a $(4,2)$ system that leads to a rotated 16-PSK constellation where each constellation point has the same amplitude. Once an average power normalization is used instead of a fixed energy constraint, the constellation plot results in a mixed pentagonal/hexagonal grid arrangement as shown in Fig.~\ref{fig:ae_consts} (c). This diagram can be compared to the 16-QAM constellation as shown in Fig.~\ref{fig:constBPSKQPSKQAM} (c).  

\begin{table}
\renewcommand{\arraystretch}{1.2} 
\centering
\caption{Layout of the autoencoder used in Figs.~\ref{fig:nnvsHamming_BPSK} (a) and (b). 
}
\label{tab:ae-layout}
\begin{tabular}{l|c|l|c}
 \multicolumn{2}{c}{\textbf{Transmitter}} & \multicolumn{2}{c}{\textbf{Receiver}}  \\ \hline \hline
 Layer & Output dimensions & Layer & Output dimensions  \\ \hline
 Input & $M$ & Input & $n$ \\
 Dense + ReLU & $M$ & Dense + ReLU & $M$ \\
 Dense + linear & $n$ & Dense + softmax & $M$\\\hline
\end{tabular}
\end{table}

In addition to promising results for the channel autoencoder implementation with simulated channels, over-the-air transmissions have also verified the feasibility of building, training, and running a complete communication system solely composed of DNNs using unsynchronized off-the-shelf software-defined radios (SDRs) and open-source deep learning software libraries \cite{DornerOTA}. Hardware implementation introduces additional challenges to the system such as the unknown channel transfer function. The autoencoder concept works when there is a differentiable mathematical expression of the channel's transfer function for each training example. A two-step training strategy is used to overcome this issue where the autoencoder is first trained with a stochastic channel model that closely approximates the real channel model. During operation time, the receiver's DNN parameters are fine-tuned using \textit{transfer learning} approach. A comparison of the BLER performance of the channel autoencoder system implemented on the SDR platform with that of a conventional communication scheme shows competitive performance close to $1$ dB without extensive hyperparameter tuning \cite{DornerOTA}.    

Transfer learning approach still provides suboptimal performance for the channel autoencoder since the channel model used during the training differs from the one experienced during operation time. A training algorithm that iterates between the supervised training of the receiver and \textit{reinforcement learning}-based training of the transmitter was developed in \cite{AoudiaAERL} for different channel models including AWGN and Rayleigh block-fading (RBF) channels. 

\subsection{Multiple Antenna Systems}\label{subsec:multiant}

MIMO wireless systems are widely used today in cellular and wireless local area network (LAN) communications. A MIMO system exploits multipath propagation through multiple antennas at the transmitter and receiver to achieve different types of gains including beamforming, spatial diversity, spatial multiplexing gains, and interference reduction. Spatial diversity is used to increase coverage and robustness by using space-time block codes (STBC) \cite{Alamouti,STBC}. Same information is precoded and transmitted in multiple time slots in this approach. Spatial multiplexing is used to increase the throughput by sending different symbols from each antenna element \cite{Telatar}, \cite{YuMac}. In a closed-loop system, the receiver performs channel estimation and sends this channel state information (CSI) back to the transmitter. The CSI is used at the transmitter to precode the signal due to interference created by the additional antenna elements operating at the same frequency. The developed MIMO schemes for both spatial diversity and multiplexing rely on analytically obtained (typically fixed) precoding and decoding schemes.

\begin{figure}[h]
\centering
\includegraphics[width=0.85\columnwidth, trim={0cm 3.2cm 0cm 2cm},clip]{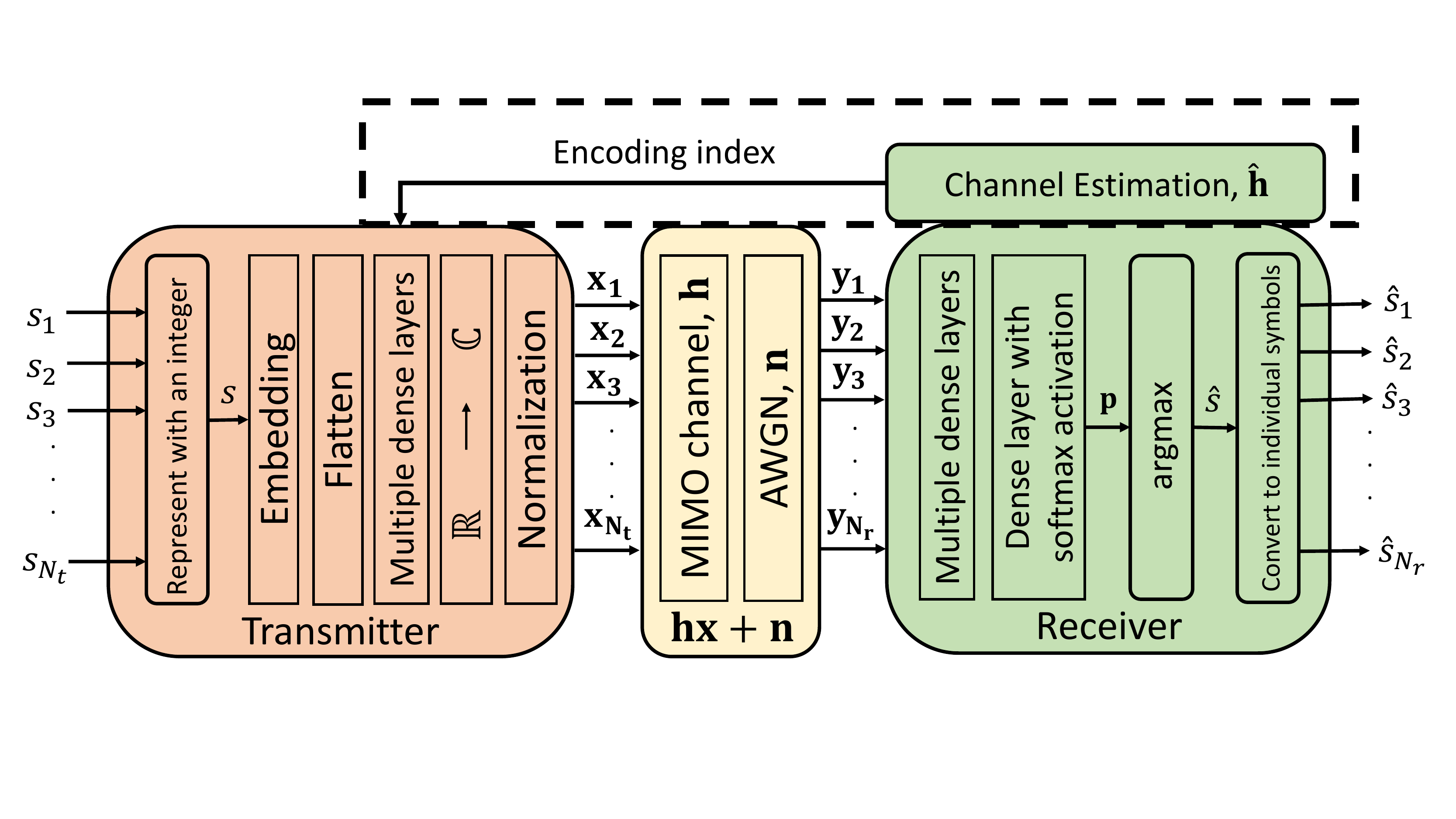}
\caption{MIMO channel autoencoder trained using a constant channel.}
\label{fig:mimocae1}
\end{figure}

Deep learning has been used for MIMO detection at the receivers to improve the performance using model-driven deep learning networks \cite{SamuelDeepMIMO,HeMIMODet,BeeryDet}. 
In Sec. \ref{subsec:singleant}, the channel autoencoder was used to train a communication system with a single antenna. The autoencoder concept is also applied to the MIMO systems where many MIMO tasks are combined into a single end-to-end encoding and decoding process which can be jointly optimized to minimize symbol error rate (SER) for specific channel conditions \cite{OSheaAllerton}. A MIMO autoencoder system with $N_t$ antennas at the transmitter and $N_r$ antennas at the receiver is shown in Fig.~\ref{fig:mimocae1}. Symbols $s_i$, $i=1,\ldots,N_t$, are inputs to the communication system. Each symbol has $k$ bits of information. By varying $k$, the data rate of the autoencoder system can be adjusted. The input symbols are combined and represented with a single integer in the range of $[0,2^{k N_t})$ as an input to the encoder (transmitter) and are encoded to form $N_t$ parallel complex transmit streams, $\mathbf{x_i}$, as output, where $i=1,\ldots,N_t$. There are different channel models developed for MIMO systems such as \cite{COST2100}. A Rayleigh fading channel is used in this example which leads to a full rank channel matrix. In this case, full benefit is achieved from the MIMO system since the received signal paths for each antenna are uncorrelated. The signal received at the decoder (receiver) can be modeled as $\mathbf{y} = \mathbf{h} \mathbf{x} + \mathbf{n}$ where $\mathbf{h}$ is an $N_r \times N_t$ channel matrix with circularly symmetric complex Gaussian entries of zero mean and unit variance, $\mathbf{x}$ is an $N_t \times 1$ vector with modulated symbols with an average power constraint of $P$ such that $\mathbb{E}\LSB\mathbf{x^*}\mathbf{x}\RSB\leq P$ where $\mathbf{x^*}$ denotes the Hermitian of $\mathbf{x}$, and  $\mathbf{n}$ is an $(N_r \times 1)$ vector which is the AWGN at the receiver with $\mathbb{E}\LSB \mathbf{n}\mathbf{n^*} \RSB=\sigma^2 \mathbf{I}_{N_r \times N_r}$. Estimated symbols $\hat{s_i}$, where $i=1,\ldots,N_r$, are the outputs. Every modulated symbol at the transmitter corresponds to a single discrete use of the channel and the communication rate of the system is $\min(N_t,N_r)\cdot k$ bits. 

The autoencoder is trained using channel realizations drawn from Rayleigh distribution.  
The transmitter communicates one out of $2^k$ possible messages from each antenna. The transmitter is designed using an FNN architecture. The input symbols go through an embedding layer followed by dense layers. Embedding layer turns positive integers to dense vectors of fixed size. The output of the embedding layer is converted to a one-dimensional tensor before going in to the dense layers using a flatten layer.  
Batch normalization \cite{Batch} is used after embedding layer and every dense layer. The output of the last dense layer is reshaped to generate complex numbers as the output; i.e., even indices as the real part and odd indices as the imaginary part. 

The transmitter has an average power constraint. The \emph{normalization} layer normalizes the transmitter output so that the average power constraint is satisfied; i.e., $E[\mathbf{x^*}\mathbf{x}]\leq P$. As in the single antenna case, the transmitter output, $\textbf{x}$ can be thought as modulated symbols as in conventional communication systems. Instead of using a known constellation scheme with linear decision regions such as BPSK or QPSK, the optimal constellation points are learned by the autoencoder system over time.   

A \emph{multiplication} layer is built to perform complex multiplication, $\mathbf{hx}$, and the \emph{noise} layer introduces noise, $\mathbf{n}$, to the autoencoder system. The input symbols and the noise change in every training instance and the noise variance $\sigma$ is adjusted at both training and test time to simulate varying levels of SNR. 

The receiver is also designed using an FNN architecture. The symbols received at the receiver, $\mathbf{y_i}$, where $i=1,\dots,N_r$, go through multiple dense layers with the last layer with softmax activation that provides a probability for each symbol with a sum equal to $1$. The codeword with the highest probability is selected as the output. 

During training, the transmitter and receiver are optimized jointly to determine the weights and biases for both of the FNNs that minimize the reconstruction loss. There are total of $2^{k N_t}$ output classes. Categorical cross-entropy loss function ($\ell_{CE}$) is used for optimization using gradient descent which is given by 

\begin{equation}
    \ell_{CE}(\boldsymbol{\theta}) = -\frac{1}{M}\sum_{i=1}^{M} \sum_{j=0}^{2^{k N_t}-1} p_{o,j}' \log(p_{o,j}),
\end{equation}
where $M$ is the mini-batch size, $\boldsymbol{\theta}$ is the set of neural network parameters, $p_{o,j}$ is the softmax layer's output probability for output class $j$ for observation $o$, and $p_{o,j}'$ is the binary indicator ($0$ or $1$) if class label $j$ is the correct classification for observation $o$. Weight updates are computed based on the loss gradient using back-propagation algorithm with Adam \cite{AdamOpt} optimizer.  In this case, a forward pass, $f(s,\boldsymbol{\theta})$, and a backward pass, 
$\frac{ \partial \ell_{CE} (\boldsymbol{\theta}) }{ \partial \boldsymbol{\theta} }$,
are iteratively computed and a weight update is given by $\delta w = -\eta \frac{\partial \ell_{CE}(\boldsymbol{\theta})}{\partial \boldsymbol{\theta}}$ with $\eta$ representing the learning rate. 

Channel estimation can be performed either using conventional or machine learning-based methods during test phase (channel estimation block in Fig.~\ref{fig:mimocae1}).  
$\mathbf{h}$ is the channel matrix and $\mathbf{\hat{h}}$ is the channel estimation at the receiver. During real-time operation, the receiver performs channel estimation and sends the index of the best encoding to the transmitter through the designated feedback channel. The cognitive transmitter will change the encoding scheme on-the-fly to minimize SER. As a result, a closed-loop system will be used during operation time as shown in Fig.~\ref{fig:mimocae1}.

Channel estimation error at the receiver leads to a sub-optimal encoding scheme to be selected both at the transmitter and the receiver, and translates to a performance loss. A minimum mean square error (MMSE) channel estimator is used at the receiver. Assuming $\mathbf{h}=\mathbf{\hat{h}}+\mathbf{\tilde{h}}$ where $\mathbf{\hat{h}}$ is the channel estimation matrix and $\mathbf{\tilde{h}}$ is the channel estimation error, the variance of $\mathbf{\tilde{h}}$ using an MMSE channel estimator is given as \cite{Hassibi}:  

\begin{equation} \label{eq:chesterror}
    \sigma_{\tilde{h}}^2 = \frac{1}{1+\frac{\rho_{\tau}}{N_t}T_{\tau}} \; ,
\end{equation}
where $\rho_{\tau}$ is the SNR during the training phase and $T_{\tau}$ is the number of training samples. (\ref{eq:chesterror}) was used in \cite{ErpekVTC, ErpekAMMO} for closed-loop MIMO systems with both channel estimation and feedback (from receiver to transmitter) to perform channel-guided precoding at the transmitter.  Different error variances are introduced to the channels (originally used for training) in test time to measure the impact of channel estimation error.  

A closed-loop MIMO system using singular value decomposition (SVD)-based precoding technique at the transmitter \cite{Telatar} is implemented as the baseline. The channel matrix, $\mathbf{h}$, can be written as $\mathbf{h}= \mathbf{U \Lambda V^*}$  where $\mathbf{U}$ and $\mathbf{V}$ are $N_r \times N_r$ and $N_t \times N_t$ unitary matrices, respectively. $\mathbf{\Lambda}$ is a diagonal matrix with the singular values of $\mathbf{h}$. To eliminate the interference at each antenna, the channel is diagonalized by precoding the symbols at the transmitter and decoding at the receiver using the CSI. In this model, the received signal is written as $\tilde{\mathbf{y}} = \mathbf{\Lambda} \tilde{\mathbf{x}} + \tilde{\mathbf{n}}$ where $\tilde{\mathbf{x}} = \mathbf{V} \mathbf{x}$, $\tilde{\mathbf{y}} = \mathbf{U^*} \mathbf{y}$ and $\tilde{\mathbf{n}} = \mathbf{U^*} \mathbf{n}$. The distribution of $\tilde{\mathbf{n}}$ is the same as $\mathbf{n}$ with $\tilde{\mathbf{n}} \sim \mathcal{N}(\mu, \, \sigma^{2} \mathbf{I}_{N_{r}})$. 

The performance of a $2\times2$ autoencoder system is evaluated and compared with the baseline performance.  
The noise variance, $\sigma^2$, is set to $1$ and $N_t = N_r$. A closed-loop system with perfect CSI (no channel estimation error) at the transmitter is assumed for the baseline simulation. QPSK modulation is used to modulate the input bits. Equal power is used at each antenna during transmission. 
A $2\times2$ autoencoder system is developed using $2$ bits per symbol to match the bit rate with the baseline. 
The FNN structures for the transmitter and receiver are  shown in Table \ref{table:paraTXRXConstant}.

\begin{table}
\caption{FNN structures used at the transmitter and receiver.}
\centering
{\small
\begin{tabular}{c|c|c|c|c}
      & \multicolumn{2}{|c|}{Transmitter} & \multicolumn{2}{c}{Receiver} \\ \hline
Layers & \# neurons & Activation function & \# neurons & Activation function\\ \hline \hline
Input & 2 &  & 4 &  \\ \hline 
 1 & 32 & ReLU & 8 & ReLU\\ \hline
 2 & 16 & ReLU & 16 & ReLU\\ \hline
 3 & 8 & ReLU & 32 & ReLU\\ \hline
 Output & 4 & Linear & 16 & Softmax \\ \hline
\end{tabular}
}
\label{table:paraTXRXConstant}
\end{table}

\begin{figure}[h]
	\centering
	\begin{tabular}{c}
		\includegraphics[width=0.45\textwidth]{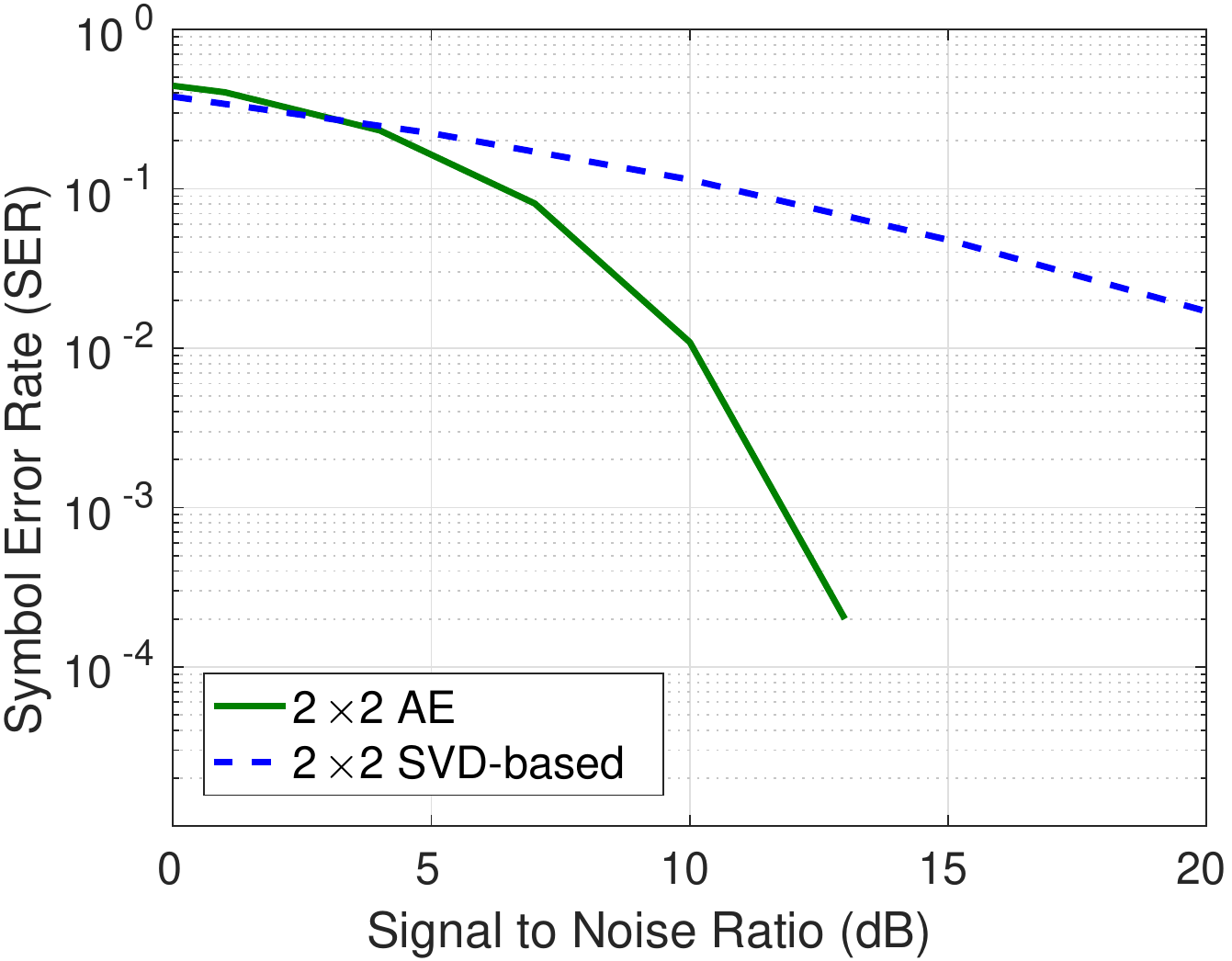} \hspace*{2em}
		\includegraphics[width=0.45\textwidth]{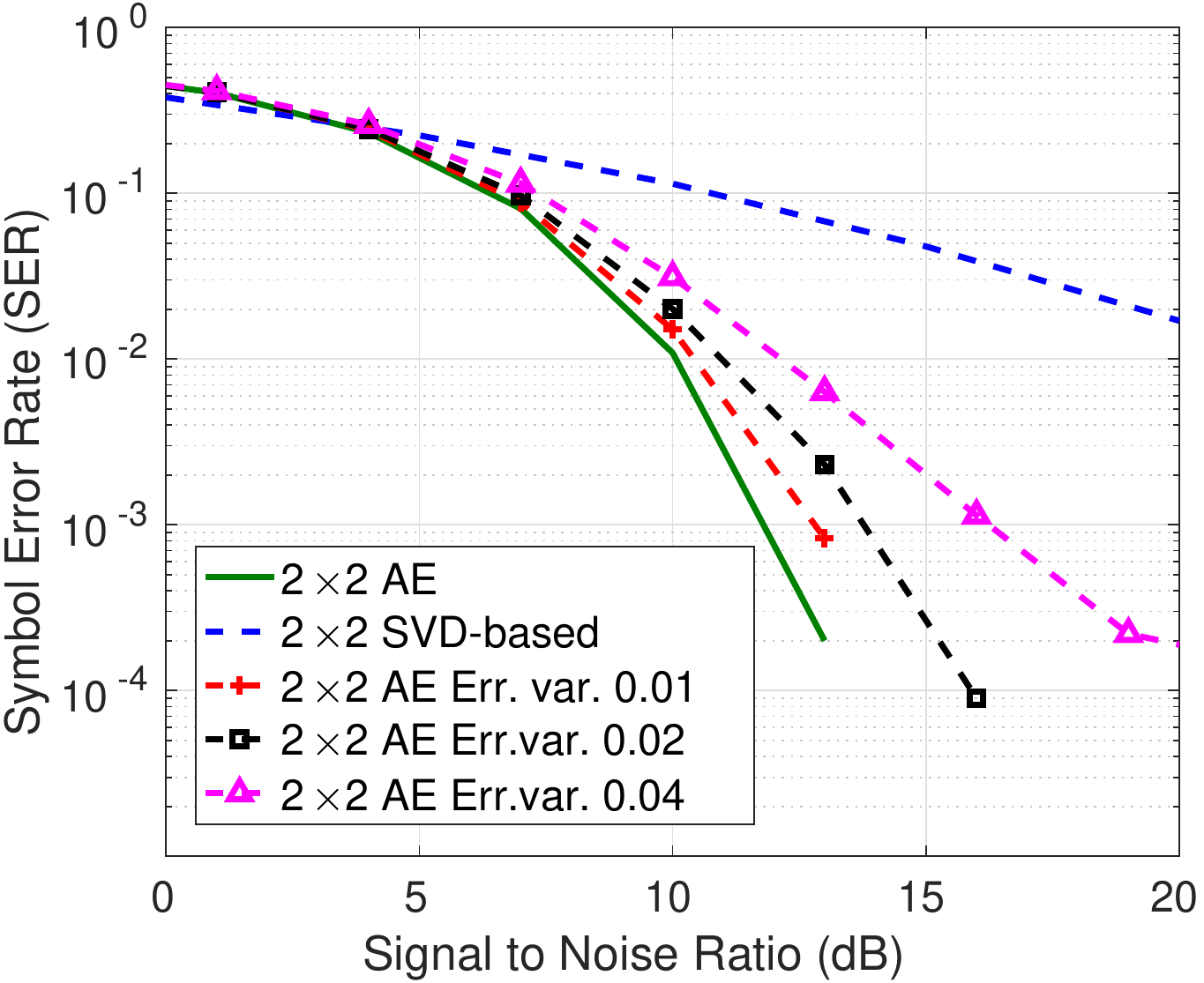}
		\\[-5pt]
		\footnotesize{\hskip9mm(a)} \hskip86.8mm \footnotesize{(b)} 
	\end{tabular}
	\caption{(a) SER performance comparison of conventional and learned $2\times2$ spatial multiplexing schemes for a constant channel with perfect CSI, (b) The effect of channel estimation error on the performance of learned $2\times2$ spatial multiplexing scheme for constant channel.}
	\label{fig:2x2serSMandError}
\end{figure}

Fig.~\ref{fig:2x2serSMandError} (a) shows the SNR vs. SER curves of the learned communication system compared to the baseline when no channel estimation error is assumed for both of the schemes. Promising results are obtained with the autoencoder approach when nonlinear constellation schemes are allowed at the transmitter. There is more than $10$ dB gain at an SER of $10^{-2}$ when the autoencoder is used.

It is assumed that the transmitter and receiver will be trained for specific channel instances and resulting neural network parameters (weights and biases) will be stored in the memory.  During operation time, the receiver will perform channel estimation and send the index of the encodings that will be used to the transmitter. There will be channel estimation error at the receiver, which increases with decreasing number of training symbols \cite{Hassibi}. Next, the performance of the developed autoencoder system when there is channel estimation error is analyzed using an MMSE channel estimator at the receiver. It is assumed that the training time increases with decreasing SNR and the system performance is analyzed when the channel estimation error variances are $0.01$, $0.02$ and $0.04$. The autoencoder system is first trained with a given channel matrix, $\mathbf{h}$. Then the output of the autoencoder architecture, weights, and biases are saved and the channel with the estimation error is provided during the operation time. Fig.~\ref{fig:2x2serSMandError} (b) shows the performance results. The autoencoder performance degrades with increasing channel estimation error, as expected. Error variance of $0.04$ is the maximum that the system can tolerate. 

\subsection{Multiple User Systems}\label{subsec:multiuser}

\begin{figure}[h]
	\centering
	\begin{tabular}{c}	
		\includegraphics[width=0.45\textwidth]{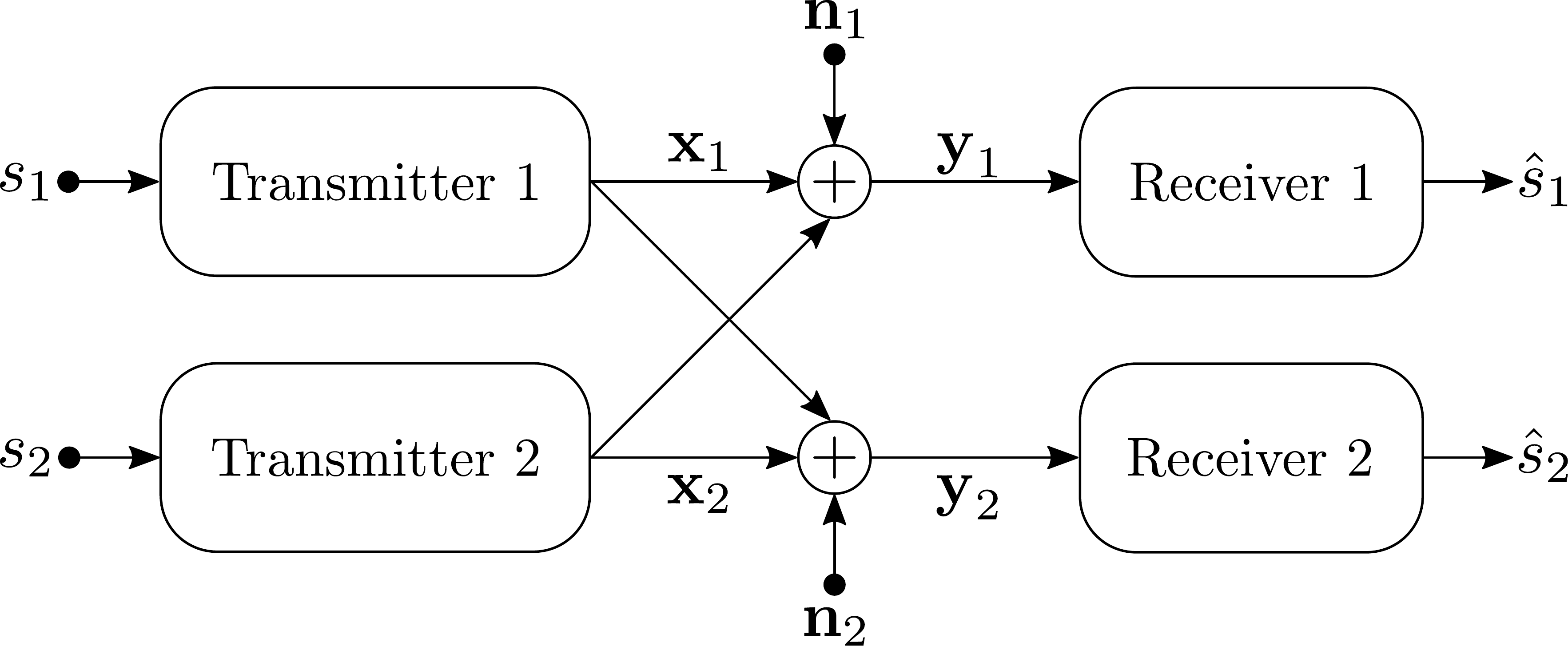}\hspace*{2em}
		\includegraphics[width=0.45\textwidth]{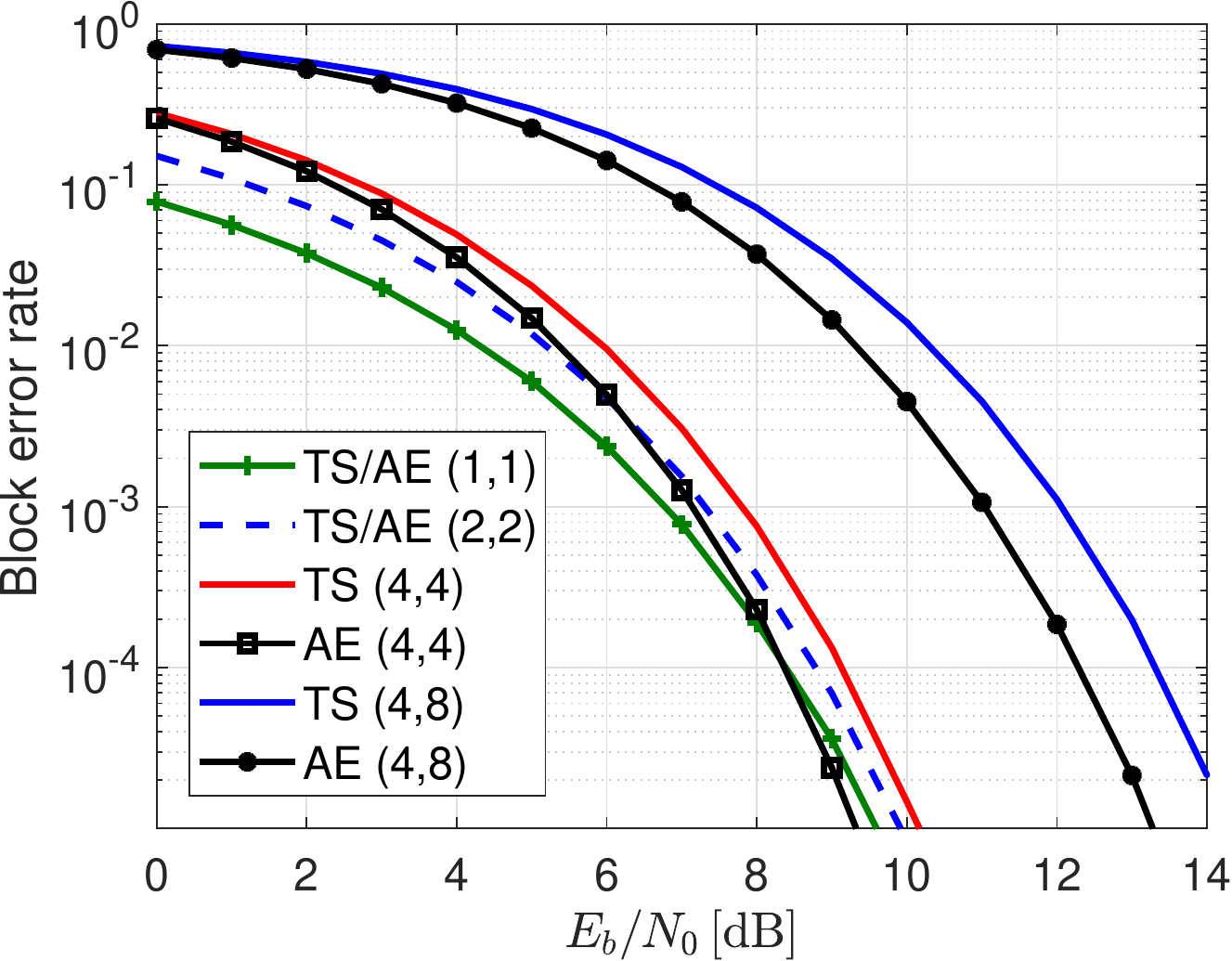}
		\\[-1pt]
		\footnotesize{\hskip8mm(a)} \hskip86.8mm \footnotesize{(b)} 
	\end{tabular}
	\caption{(a) The two-user interference channel seen as a combination of two interfering autoencoders (AEs) that try to reconstruct their respective messages, (b) \Ac{BLER} versus $E_b/N_0$ for the two-user interference channel achieved by the autoencoder and $2^{2k/n}$-\Ac{QAM} time-sharing (TS) for different parameters $(n,k)$.}
	\label{fig:ae_ic_ic_vs_timesharing}
\end{figure}

The autoencoder concept described in Sec.~\ref{subsec:singleant} was extended to multiple transmitters and receivers that operate at the same frequency for single antenna systems in \cite{OSheaTCCN} and for multiple antenna systems in \cite{ErpekICC}. A two-user \Ac{AWGN} interference channel was considered in \cite{OSheaTCCN} as shown in Fig.~\ref{fig:ae_ic_ic_vs_timesharing} (a).

Transmitter~1 wants to communicate message $s_1\in\mathbb{M}$ to Receiver~1 and simultaneously, Transmitter~2 wants to communicate message $s_2\in\mathbb{M}$ to Receiver~2. Extensions to $K$ users with possibly different rates and other channel types are straightforward. Both transmitter-receiver pairs are implemented as FNNs. The encoder and decoder architectures are the same as described in Sec.~\ref{subsec:singleant}. However, the transmitted messages interfere at the receivers in this case. The signal received at each receiver is given by
\begin{equation}
 \yv_1 = \xv_1 + \xv_2 + \nv_1, \quad \yv_2 = \xv_2 + \xv_1 + \nv_2,
\end{equation}
where $\xv_1, \xv_2 \in \mathbb{C}^n$ are the transmitted messages and $\nv_1,\nv_2\sim\Cc\Nc(0,\beta\Id_n)$ is Gaussian noise. No fading is assumed in this scenario; i.e., $\mathbf{h}$ values are set to $1$ for each link. The individual cross-entropy loss functions of the first and second transmitter-receiver pairs are $l_1 = -\log\LB\LSB\hat{\sv}_1\RSB_{s_1}\RB$ and $l_2 = -\log\LB\LSB\hat{\sv}_2\RSB_{s_2}\RB$ for the first and second autoencoder, respectively. 

$\tilde{L}_1(\thetav_t)$, and $\tilde{L}_2(\thetav_t)$ correspond to the associated losses for mini-batch $t$. For joint training, dynamic weights $\alpha_t$ are adapted for each mini-batch $t$ as

\begin{align}
\alpha_{t+1} = \frac{\tilde{L}_1(\thetav_t)}{\tilde{L}_1(\thetav_t) + \tilde{L}_2(\thetav_t)},\quad t>0 \; ,
\end{align}
where $\alpha_0=0.5$. Thus, the smaller $\tilde{L}_1(\thetav_t)$ is compared to $\tilde{L}_2(\thetav_t)$, the smaller is its weight $\alpha_{t+1}$ for the next mini-batch.  

Fig.~\ref{fig:ae_ic_ic_vs_timesharing} (b) shows the \Ac{BLER} of one of the autoencoders as a function of $E_b/N_0$ for the sets of parameters $(n,k)=\{(1,1), (2,2), (4,4), (4,8)\}$. The DNN architecture for both autoencoders is the same as that provided in Table~\ref{tab:ae-layout} by replacing $n$ by $2n$. An average power constraint is used to be competitive with higher-order modulation schemes; i.e., allow varying amplitude in the constellation points for increasing data rate. As a baseline, uncoded $2^{2k/n}$- \Ac{QAM} (which has the same rate when used together with time-sharing between both transmitters) is considered. For $(1,1)$, $(2,2)$, and $(4,4)$, each transmitter sends a 4-\Ac{QAM} (i.e., \Ac{QPSK}) symbol on every other channel use. For $(4,8)$, 16-\Ac{QAM} is used instead. While the autoencoder and time-sharing have identical \Ac{BLER} for $(1,1)$ and $(2,2)$, the former achieves substantial gains of around $0.7$ dB for $(4,4)$ and $1$ dB for $(4,8)$ at a \Ac{BLER} of $10^{-3}$.

\begin{figure}
\centering
\begin{tabular}{l}
  \hskip-3mm\includegraphics[width=0.15\textwidth]{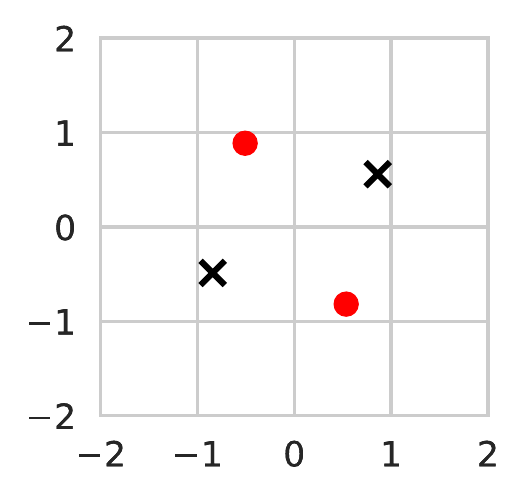}        \hskip7mm\includegraphics[width=0.15\textwidth]{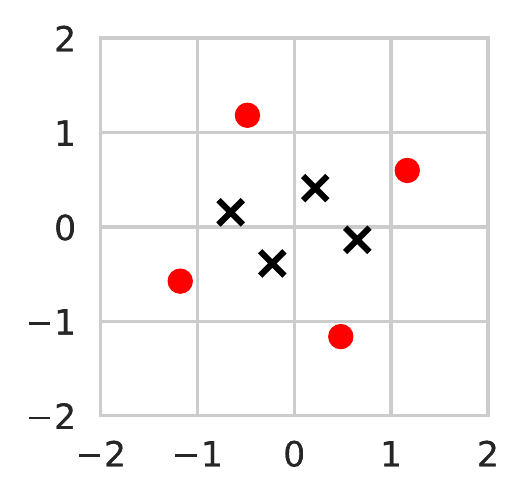} 
   \hskip-1mm\includegraphics[width=0.15\textwidth]{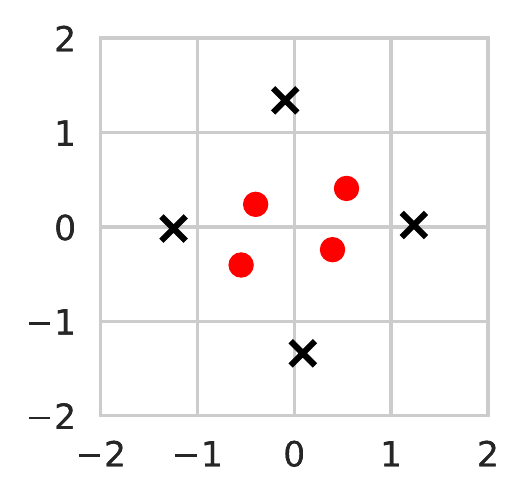} \\[-5pt]   
 \hskip11mm \footnotesize{(a)} \hskip45mm  \footnotesize{(b)}\\[-8pt] 
 \\
\hskip-3mm\includegraphics[width=0.15\textwidth]{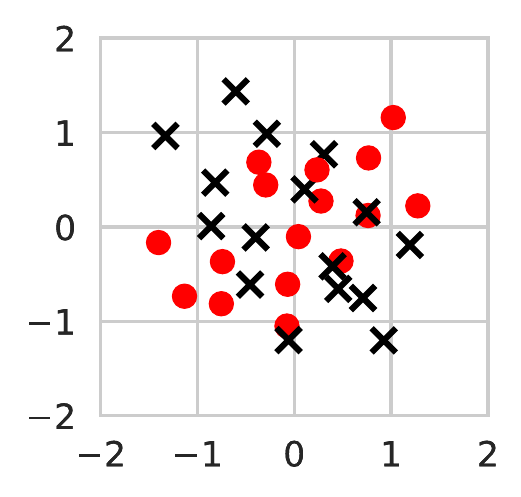}         \hskip-1mm\includegraphics[width=0.15\textwidth]{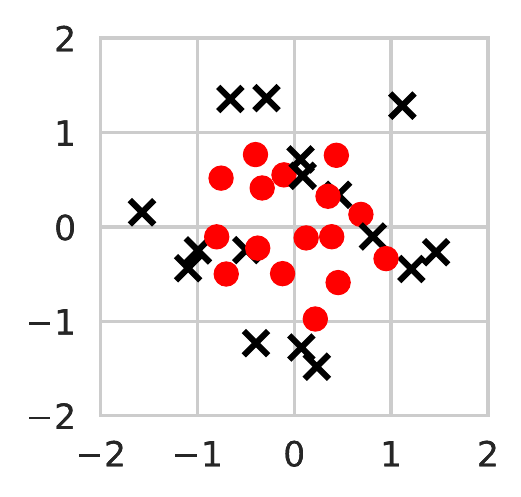} 
    \hskip-1mm\includegraphics[width=0.15\textwidth]{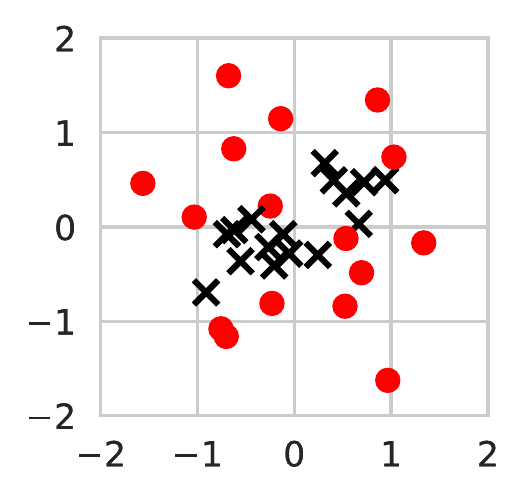}
    \hskip-1mm\includegraphics[width=0.15\textwidth]{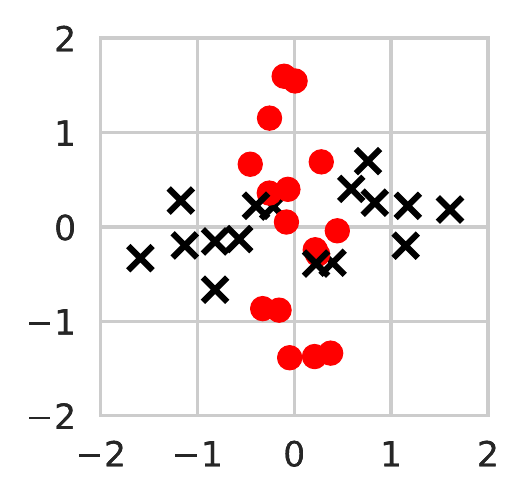}\\[-5pt]
     \hskip53mm \footnotesize{(c)}
      \\
\hskip-3mm\includegraphics[width=0.15\textwidth]{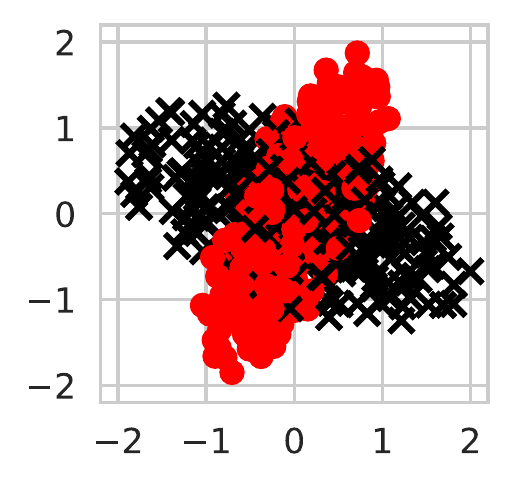}         \hskip-1mm\includegraphics[width=0.15\textwidth]{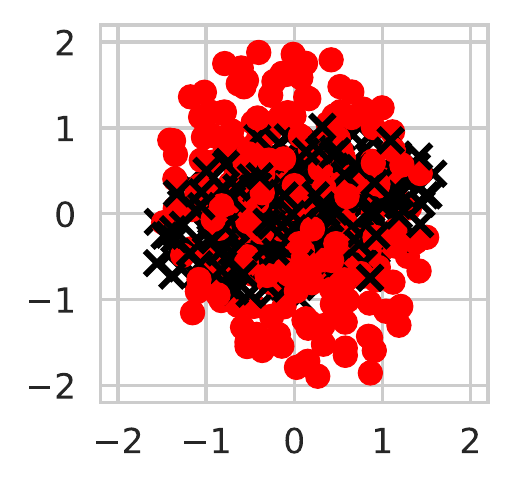} 
    \hskip-1mm\includegraphics[width=0.15\textwidth]{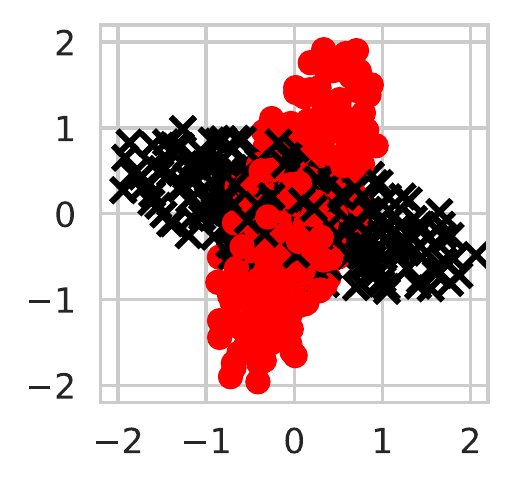}
    \hskip-1mm\includegraphics[width=0.15\textwidth]{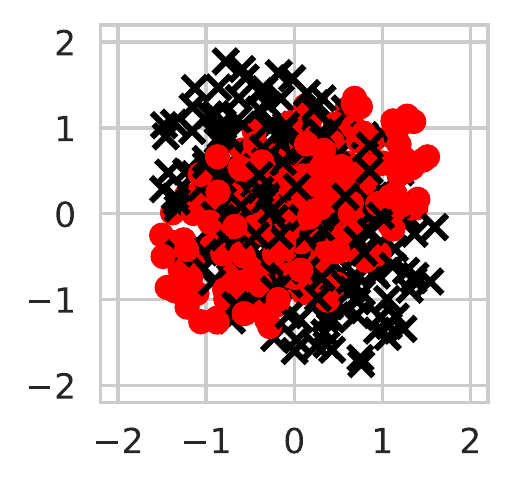}\\[-5pt]
     \hskip53mm \footnotesize{(d)}
\end{tabular}
\caption{Learned constellations for the two-user interference channel with parameters (a) $(1,1)$, (b) $(2,2)$, (c) $(4,4)$, and (d) $(4,8)$. The constellation points of Transmitters~1 and 2 are represented by red dots and black crosses, respectively \cite{OSheaTCCN}.}\label{fig:ic_constellations}
\end{figure}

The learned message representations at each receiver are shown in Fig.~\ref{fig:ic_constellations}. For $(1,1)$, the transmitters have learned to use \Ac{BPSK}-like constellations (see Fig.~\ref{fig:constBPSKQPSKQAM} (a)) in orthogonal directions (with an arbitrary rotation around the origin). This achieves the same performance as \Ac{QPSK} with time-sharing. However, for $(2,2)$, the learned constellations are not orthogonal anymore and can be interpreted as some form of superposition coding. For the first symbol, Transmitter~1 uses high power and Transmitter~2 uses low power. For the second symbol, the roles are changed. For $(4,4)$ and $(4,8)$, the constellations are more difficult to interpret, but it can be seen that the constellations of both transmitters resemble ellipses with orthogonal major axes and varying focal distances. This effect is more visible for $(4,8)$ than for $(4,4)$ because of the increased number of constellation points. 

\textbf{Take-away:} This section showed that deep learning-based autoenconder can be effectively used to develop transmitter (modulation and coding) and receiver (demodulation and decoding) functions jointly by combating channel impairments and optimizing end-to-end communication performance in terms of error rates. This approach applies to single, multiple antenna, and multiuser systems. 

\section{Deep Learning for Spectrum Situation Awareness} \label{sec:sitawar}
Cognitive radio has emerged as a programmable radio that aims to learn from wireless communication data and adapt to spectrum dynamics. For that purpose, cognitive radio senses its operational radio frequency (RF) environment and adjusts its operating parameters (e.g., frequency, power, and rate) dynamically and autonomously to modify system operation and improve its performance, such as maximizing throughput, mitigating interference, facilitating interoperability, or accessing spectrum as a secondary user \cite{FCC}. 

Channel modeling is important while developing algorithms to enable cognitive capabilities and evaluating the performance of the communication systems. Most signal processing algorithms applied to wireless communications assume compact mathematically convenient channel models such as AWGN, Rayleigh, or Rician fading channel (or fixed delay/Doppler profiles consisting of Rayleigh fading taps).  These existing channel models generally parameterize channel effects in a relatively rigid way which does not consider the exact statistics of deployment scenarios.  Furthermore, practical systems often involve many hardware imperfections and non-linearities that are not captured by these existing channel models \cite{OSheaTCCN}. Channel estimation is also an important task for a communication system to recover and equalize the received signal (reversing the channel effects). A known training sequence is often transmitted at the transmitter and the receiver typically uses methods such as maximum likelihood or MMSE channel estimation techniques, derived under compact mathematical channel models, to estimate the channel, e.g., MMSE estimator is applied in (\ref{eq:chesterror}) for channel estimation in Sec.~\ref{subsec:multiant}.  

To support situational awareness, it is important for cognitive radios to quickly and accurately perform signal detection and classification tasks across a wide range of phenomena. One example is the DSA application where there are primary (legacy) and secondary (cognitive) users. Secondary users use the spectrum in an opportunistic manner by avoiding or limiting their destructive levels of interference to the primary users in a given frequency band. Therefore, secondary users need to detect and classify the signals received during spectrum sensing reliably to identify whether there is any primary user activity, other secondary users, or vacant spectrum opportunities. Conventional signal detection and classification algorithms aim to capture specific signal features (i.e., expert features) such as cyclostationary features and are typically developed to achieve performance goals such as detection against specific signal types and under specific channel model assumptions (e.g., AWGN). 
Therefore, these conventional algorithms often lack the ability to generalize to different signal types and channel conditions, while deep learning can capture and adapt its operation to raw and dynamic spectrum data of a wide variety of signal signatures and channel effects (that feature-based machine learning algorithms may struggle to capture).

Deep learning approaches have been used to address the challenges associated with both channel modeling and estimation as well as signal detection and classification tasks. In the following subsections we first describe how channel modeling and estimation can be performed using deep learning methods. Next, we describe the CNN architectures that are used for signal detection and modulation classification. Finally, we describe how to use GANs to augment training data in spectrum sensing applications.    

\subsection{Channel Modeling and Estimation}\label{subsec:chest}

The performance of communication systems can often benefit from being optimized for specific scenarios which exhibit structured channel effects such as hardware responses, interference, distortion, multi-path and noise effects beyond simplified analytic models or distributions. Moreover, the channel autoencoder systems described in Sec.~\ref{sec:endtoend} requires the statistical model for the channel be as close as possible to what the operational system will experience during training in order to achieve optimal performance (i.e., the phenomena during training should accurately match the phenomena during deployment). However, accurately capturing all these effects in a closed-form analytical model is a challenging (and often infeasible) task. As a result, the channel is often represented using simplified models without taking real-world complexities into account.  Recently, model-free approaches where the channel response is learned from data are proposed for real-time channel modeling using deep learning techniques. In particular, stochastic channel response functions are approximated using GANs \cite{OSheaVoid,LiGAN}, variational GANs \cite{OSheaChannelModel}, reinforcement learning and sampling approach \cite{GrathwohlRL}, stochastic perturbation techniques \cite{raj2018backpropagating}, and reinforcement learning policy gradient methods \cite{aoudia2018end}.

\begin{figure}[!ht]
    \centering
    \includegraphics[width=0.7\textwidth, trim={0cm 0cm 0cm 0cm},clip]{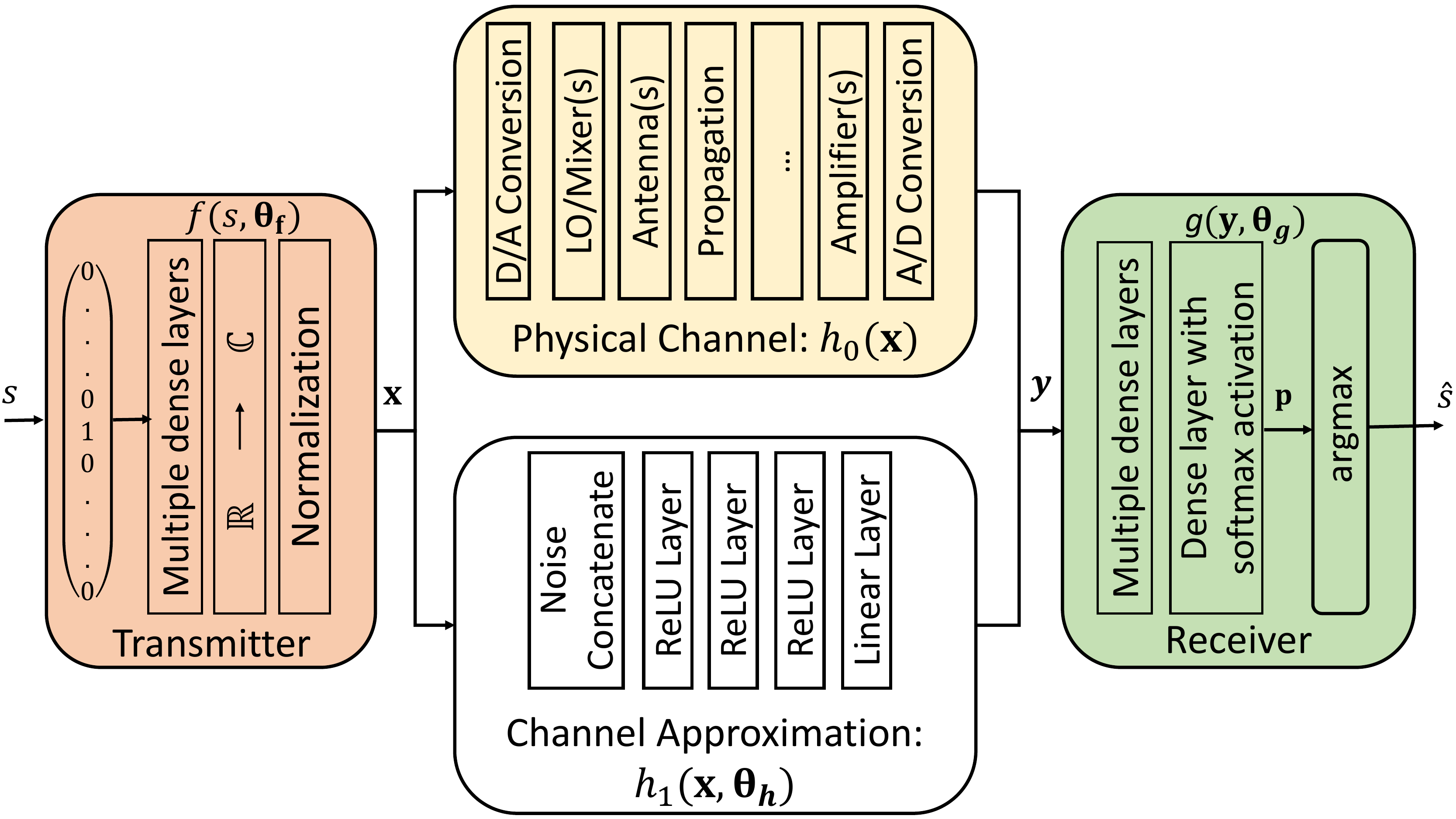}
    \caption{A GAN for learning a communication system over a physical channel with no-closed form model.}
    \label{fig:GANFigure}
\end{figure}

GANs \cite{goodfellow2014generative} have been successfully used for a number of applications such as generating fake images (e.g., faces, cats) to confuse image recognition systems. Recently, GANs have also been used in a wide range of applications such as audio generation, approximation of difficult distributions, and even the (human-guided) generation of novel art.  Building upon this same idea, the GAN was applied to approximate the response of the channel in any arbitrary communication system in \cite{OSheaVoid} and the resulting system was generally called a \textit{Communications GAN}. The block diagram of the Communications GAN that learns a communication system over a physical channel with no closed-form model or expression is shown in Fig.~\ref{fig:GANFigure}.  

As opposed to the original autoencoder shown in Fig.~\ref{fig:ae_awgn}, a channel model with an analytic expression is not included in the autoencoder in Fig.~\ref{fig:GANFigure}. Two forms of the channel $h(\mathbf{x})$ are included instead to encompass modeling of any black-box channel transform where $\mathbf{x}$ is the transmitter output: $h_0(\mathbf{x})$ is a real-world physical measurement of the response of a communication system comprising a transmitter, a receiver, and a channel and $h_1(\mathbf{x},\mathbf{\theta_h})$ is a non-linear DNN which seeks to mimic the channel response of $h_0$ synthetically, and is differentiable. $\mathbf{\theta_h}$ is the channel approximation of neural network parameters. During training, an iterative approach is used to reach an optimized solution, cycling between competing training objectives, updating weights for each network during the appropriate stage with manually tuned learning rates and relatively small networks for $f$, $g$, and $h$, and employing several fully connected ReLU layers for each. The physical channel $h_0(\mathbf{x})$ was implemented using an SDR (Universal Software Radio Peripheral, USRP B210 \cite{ettus2009universal}), for over-the-air transmission tests. It was shown that an effective autoencoder-based communication system with robust performance can be learned by using an adversarial approach to approximate channel functions for arbitrary communications channel. This approach eliminates the need for a closed-form channel model reducing the need for assumptions on the form it takes.

The channel network $\mathbf{y} = h(\mathbf{x})$ is treated as a stochastic function approximation and the accuracy of the resulting conditional probability distribution $p(\mathbf{y}|\mathbf{x})$ is optimized in \cite{OSheaChannelModel}.   
The channel approximation network $\mathbf{\hat{y}} = h(\mathbf{x},\mathbf{\theta_h})$ is considered to be a conditional probability distribution, $p(\mathbf{\hat{y}}|\mathbf{x})$ and the distance between the conditional probability distributions $p(\mathbf{y}|\mathbf{x})$ and $p(\mathbf{\hat{y}}|\mathbf{x})$ resulting from the measurement and from the variational channel approximation network are minimized.  As in \cite{goodfellow2014generative}, the parameters of each network are minimized using the two stochastic gradients given in (\ref{eq:bceloss1}) and (\ref{eq:bceloss2}).
\begin{equation}
    \label{eq:bceloss1}
    \nabla_{\mathbf{\theta_D}} \frac{1}{N} \sum_{i=0}^{N} \left[ \text{log} \left( D(x_i,y_i,\mathbf{\theta_D}) \right) + \text{log}\left( 1 - D(x_i,h(x_i,\mathbf{\theta_h}),\mathbf{\theta_D})  \right) \right],
\end{equation}
\begin{equation}
    \label{eq:bceloss2}
    \nabla_{\mathbf{\theta_h}} \frac{1}{N} \sum_{i=0}^{N} \text{log} \left( 1 - D(x_i,h(x_i,\mathbf{\theta_h}), \mathbf{\theta_D}) \right).
\end{equation}
A new discriminative network $D(x_i,y_i,\mathbf{\theta_D})$ is introduced to classify between real samples, $\mathbf{y}$, and synthetic samples, $\mathbf{\hat{y}}$, from the channel given its input, $\mathbf{x}$. $\mathbf{\theta_D}$ is the discriminative network parameters. $h(\mathbf{x},\mathbf{\theta_h})$ takes the place of the generative network, $G(z)$, where $\mathbf{x}$ reflects conditional transmitted symbols/samples. $N$ is the number of samples. Additional stochasticity in the function is introduced through variational layers.  Furthermore, training such an arrangement using the improved Wasserstein GAN approach with gradient penalty (WGAN-GP) \cite{wgangp} allows convergence with minimal tuning.

\begin{figure}[h]
	\centering
	\begin{tabular}{c}
		\includegraphics[width=0.4\textwidth, trim={3cm 10cm 15cm 0cm},clip]{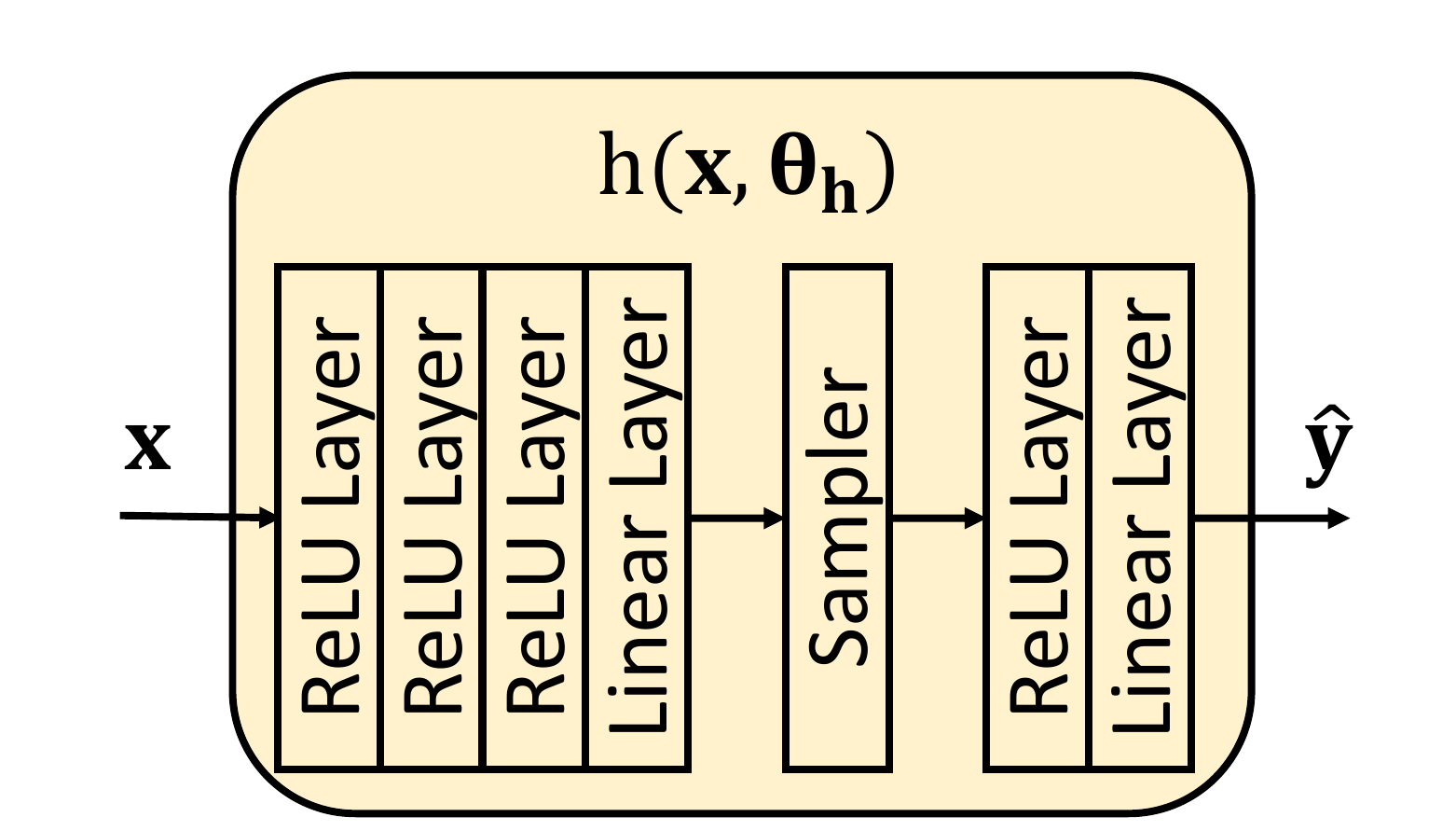} \hspace*{1em}
		\includegraphics[width=0.5\textwidth]{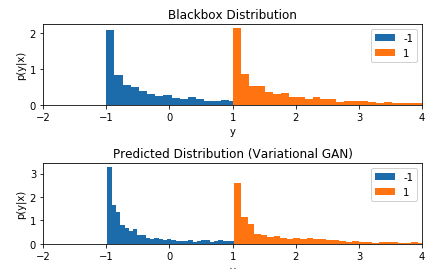}
		\\[-1pt]
		\footnotesize{\hskip0mm(a)} \hskip86.8mm \footnotesize{(b)}
	\end{tabular}
	\caption{(a) Variational architecture for the stochastic channel approximation network (conditional generator), (b) Learned one-dimensional distributions of conditional density on non-Gaussian (Chi-Squared) channel effects using variational GAN training \cite{OSheaChannelModel}.}
	\label{fig:varchan_distchi}
\end{figure}

Adam \cite{AdamOpt} optimizer is used with a learning rate between $10^{-4}$ and $5\times10^{-4}$ to iteratively update the network parameters. The variational architecture for the stochastic channel approximation network is shown in Fig.~\ref{fig:varchan_distchi} (a).

For performance evaluation, a communication system that transmits $1$ bit/symbol is considered. A Chi-squared distributed channel model is assumed to explore a more uncommon channel scenario. The measured and approximated conditional distributions from the black box channel model are shown in Fig.~\ref{fig:varchan_distchi} (b). There is some difference between the original distribution and its approximation, resulting partially from its representation as a mixture of Gaussian latent variables; however, this can be alleviated by choosing different sampling distributions and by increasing the dimensions of the latent space (at the cost of increased model complexity).

\begin{figure}[!ht]
    \centering
    \includegraphics[width=0.6\textwidth]{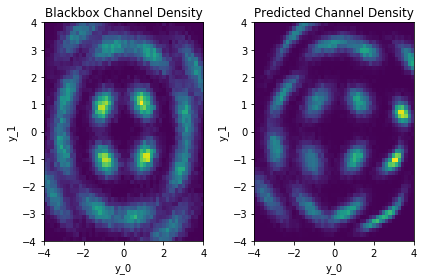}
    \caption{Learned two-dimensional distributions of received 16-QAM constellation non-linear channel effects using variational GAN \cite{OSheaChannelModel}.}
    \label{fig:dist2d2}
\end{figure}

This approach can also capture more complex distributions such as the channel responses of cascades of stochastic effects by jointly approximating the aggregate distribution with the network. Consider a 16-QAM system that includes AWGN effects along with phase noise, phase offset, and non-linear AM/AM and AM/PM distortion effects introduced by a hardware amplifier model. Fig.~\ref{fig:dist2d2} illustrates the marginalized $p(\mathbf{x})$ distribution for both the measured version of the received signal, and the approximated version of the distribution when a stochastic channel approximation model is learned with variational GANs.  It is observed that each constellation point's distribution, circumferential elongation of these distributions due to phase noise at higher amplitudes, and generally the first order approximation of the distribution are learned successfully. 

On the receiver side, typically synchronization is performed on the signal (timing estimation, frequency offset estimation, etc.) before performing additional signal processing steps for conventional communication systems (e.g., symbol detection). Synchronization typically estimates these time, frequency, phase, and rate errors in the received data and corrects for them to create a normalized version of the signal. Learned communication systems described in Sec.~\ref{sec:endtoend} can in some instances perform implicit synchronization and channel estimation since hardware and channel impairments such as synchronization offsets can be included during training. From a learning perspective, we can treat these corrections as transforms, leveraging expert knowledge about the transforms to simplify the end-to-end task, but still allowing the estimators to be fully learned.  This approach of radio transformer networks (RTNs), as explored in both of \cite{OSheaRTN,OSheaTCCN}, are shown to reduce training time and complexity and improve generalization by leveraging domain knowledge.  
 
These offset effects exist in any real system containing transmitters and receivers whose oscillators and clocks are not locked together. 

Timing and symbol-rate recovery processes involve the estimation and re-sampling of the input signal at correct timing offsets and sampling increments, which has a direct analogue to the extraction of visual pixels at the correct offset, shift or scale (e.g., applying the correct Affine transformation) in computer vision using transformer networks. The input data can be represented as a two-dimensional input, with the rows containing in-phase (I) and quadrature (Q) samples and N columns containing samples in time. A full 2D Affine transformation allows for translation, rotation, and scaling in 2D given by a $2\times3$ element parameter vector. To restrict this to 1D translation and scaling in the time dimension, the mask in (\ref{eq:timetransform}) is introduced such that a normal 2D Affine transform implementation may be used from the image domain. $\theta_0$, $\theta_1$, and $\theta_2$ are the remaining unmasked parameters for the 1D Affine transform:

\begin{equation} \label{eq:timetransform}
  \left[ {\begin{array}{ccc}
   \theta_0 & 0 & \theta_2 \\
   0 & \theta_1 & 0 \\
  \end{array} } \right]
\end{equation} 

Phase and frequency offset recovery tasks do not have an immediate analogue in the vision domain. However, a simple signal processing transform can be applied to accomplish these.  The input signal is mixed with a complex sinusoid with phase and frequency as defined by two new unknown parameters as shown in (\ref{eq:phasetransform}).

\begin{equation} \label{eq:phasetransform}
y_n = x_n \: \mathrm{e}^{\:j(n\theta_3+\theta_4)} 
\end{equation}
This transform can be directly implemented as a new layer in Keras \cite{Keras}, cascaded before the Affine transform module for timing and symbol-rate recovery. 

\begin{figure}[!ht]
    \centering
    \includegraphics[width=0.95\textwidth]{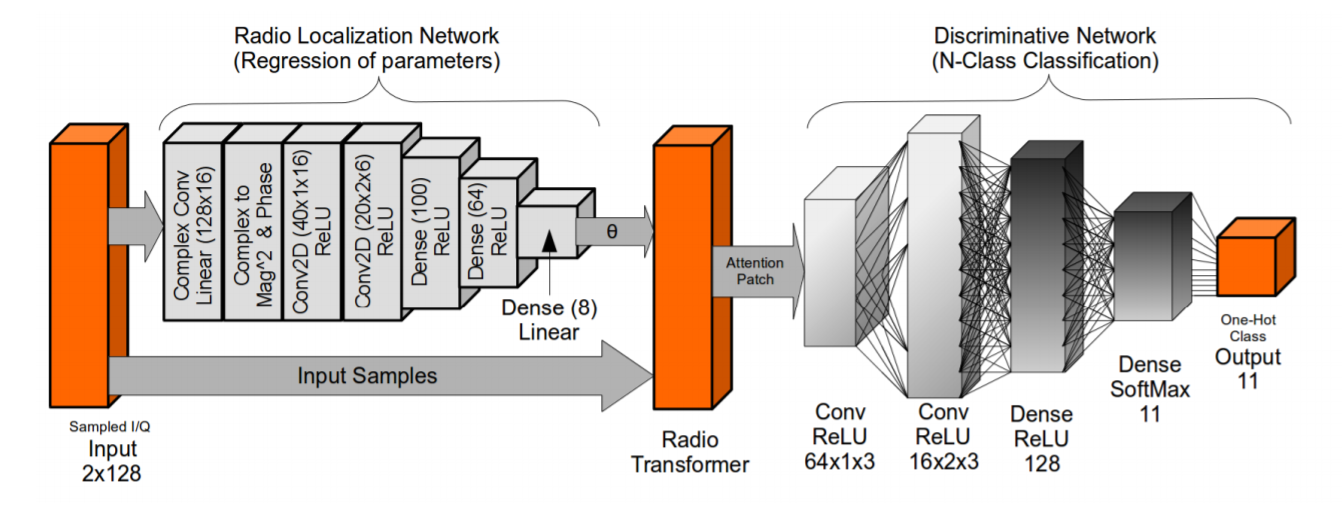}
    \caption{RTN architecture \cite{OSheaRTN}.}
    \label{fig:RTN}
\end{figure}

The task of synchronization then becomes the task of parameter estimation of $\theta_i$ values passed into the transformer modules. Domain appropriate layers are used to assist in estimation of these parameters, namely, complex convolutional 1D layer and complex to power and phase layers. Although many architectures are possible, both the complex convolution operation and the differentiable Cartesian to Polar operation are used to simplify the learning task. Fig.~\ref{fig:RTN} shows one example of an RTN architecture. A dropout rate such as $0.5$ can used between layers to prevent over-fitting, and Adam \cite{AdamOpt} SGD can be used to optimize network parameters on the training set, in this case with batch size $1024$, and learning rate $0.001$. 

\begin{figure}[!ht]
    \centering
    \includegraphics[width=0.7\textwidth]{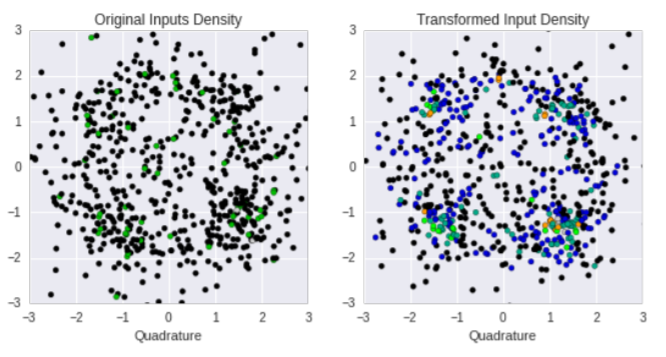}
    \caption{Density plots of the pre- and post-transformed input constellations \cite{OSheaRTN}.}
    \label{fig:RTNPerf}
\end{figure}

The density plots for pre- and post-transformed input constellations are shown in Fig.~\ref{fig:RTNPerf}. When the constellation density for $50$ test examples over a range of $20$ time samples are observed, the density starts to form around the constellation points after using the radio attention model.

In both \cite{OSheaTCCN,AoudiaAERL}, Rayleigh block fading channel is considered as the channel and RTNs are used for channel estimation. Then the  received  signal  is  divided  by  the learned channel response to equalize the input signal, which leads to improved SER performance, providing a more quantitative study of the RTN efficacy.  

The described channel modeling approaches may be used broadly for enhanced optimization, test, and measurement of communication systems and specifically to provide effective model-free methods for various wireless tasks such as channel learning in autoencoder-based communications (see Sec. \ref{sec:endtoend}) and signal classification (see Sec. \ref{subsec:detectandclassify}). Moreover, the developed RTN models can be used to extract the channel features, similar to channel estimation in conventional systems, and perform equalization by using a transformation layer which allows for imparting of expert knowledge without over-specifying learned models (e.g., writing estimators for specific protocols or references).

\subsection{Signal Detection and Modulation Classification}\label{subsec:detectandclassify}

Signal detection and classification functionalities define the ability of a wireless communication system to accurately build and maintain an up-to-date view of their current operating environment. Detecting and coexisting with other users of the spectrum, detecting and isolating sources of interference, flagging significant spectral events, or identifying spectral vacancies within the radio spectrum rely on signal detection and classification. The probability of detection is proportional to the SNR at the receiver. Traditionally, specific signal detectors are needed for each waveform, developed based on its analytic properties, resulting in systems which can be difficult to develop and deploy robustly in real-world wireless applications largely due to their over-specificity, complexity, or sub-optimal performance in real world conditions. \cite{OSheaAsilomar}.

The RF spectrum is shared with many different signal types ranging from TV broadcast to radar. Signal detection and classification tasks are particularly challenging in the presence of multiple waveforms operating at the same frequency and at low SNR. Conventional signal detection and classification methods can be categorized as:
\begin{itemize}
    \item \textit{General methods}: These methods do not require any prior information on the signal types. They detect multiple signal types; however, their constant false alarm rate (CFAR) performance is relatively poor. Energy detector \cite{AkyildizCRSurvey} is an example of detectors which do not require prior information. These type of detectors can be easily cast into convenient probabilistic form for analysis, but they are severely constrained in their abilities to leverage  additional information about signal context or structure to improve performance.  
    \item \textit{Specialized methods}: These methods provide sensitive detectors for specific signal types. The detection and classification methods are developed using specific features of the signal of interest. Matched filters and cyclostationary signal detectors \cite{AkyildizCRSurvey} are examples to this type. These methods are often not scalable since a new type of classifier is required for each new waveform.  
\end{itemize}

A new class of deep learning-based radio waveform detectors that leverages the powerful new techniques developed in computer vision, especially convolutional feature learning, holds the potential to improve the signal detection and classification performance of practical systems by generalizing well and remaining sensitive to very low power signals \cite{OSheaAsilomar}. A strong analogy of this task exists in computer vision with object identification and localization tasks. Recent object detection and localization approaches associate specific object classes with bounding box labels within the image. A similar approach was followed in \cite{OSheaEusipco}, where the RF spectrum is represented as an image and CNNs are used to detect, localize and identify radio transmissions within wide-band time-frequency power spectrograms using feature learning on 2D images.  

\begin{figure}
  \centering
      \includegraphics[width=0.8\textwidth]{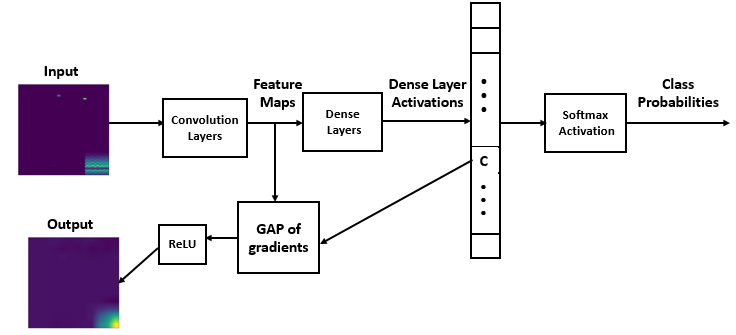}
  \caption{Block Diagram of Grad-CAM \cite{selvaraju2016grad}.}
  \label{fig:block_diag}  
\end{figure}

Gradient-weighted Class Activation Mapping (Grad-CAM) uses the gradients of any target concept flowing into the final convolutional layer to produce a coarse localization map highlighting the important regions in the image that aids to predict the concept \cite{selvaraju2016grad}. Grad-CAM is used to perform the spectral event localization in \cite{OSheaEusipco}. Fig.~\ref{fig:block_diag} shows the block-diagram of the Grad-CAM, which is used for spectral event localization. The gradient of activation score $y^C$ (instead of the class probability) is calculated with respect to all the feature maps of a given convolution layer based on the provided input label $C$. The global average pooling \cite{LinCY13} of the gradients gives the corresponding weight associated with the feature map. Finally, the weighted sum of the feature maps is passed through an element-wise ReLU unit to get the class activation map.

\begin{figure*}[!ht]
  \includegraphics[width=\textwidth,height=8cm]{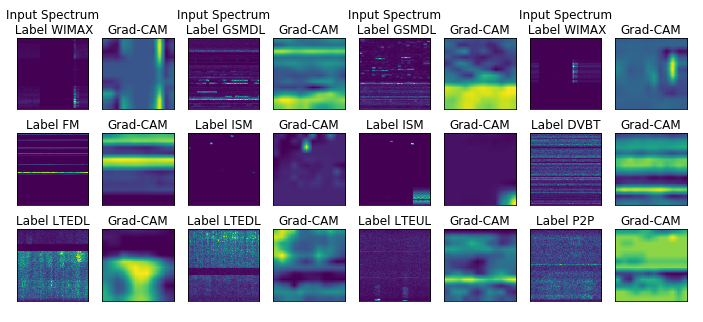} 
  \caption{GradCAM based activation maps and corresponding input spectrograms for 12 test examples from the dataset \cite{OSheaEusipco}.}
  \label{fig:grad_cam}  
\end{figure*}

To demonstrate the performance in this work, a dataset was collected in $13$ different frequency bands using a USRP B205 transceiver at eight different locations across five distinct cities and across a range of different bands and traffic patterns.  Signal types in the dataset include GSM, LTE, ISM, TV, and FM among others. Spectrogram plots shown in Fig.~\ref{fig:grad_cam}, labeled as \textit{input spectrum}, are generated using the collected data to show the signal strength over time and frequency. The x-axis shows the time and the y-axis shows the signal frequency. These images are used as an input to the CNN architecture. The Grad-CAM implementation results are also shown in Fig.~\ref{fig:grad_cam}. A hot region of activation is observed on top of the signal bursts, as expected. The trained feature objective was to classify the band instead of activating all instances of a certain emission type since the labels for every signal activity in a band are not provided; i.e., each spectrogram is assigned only one label even though there may be some other narrow band signals in the same spectrogram. For this reason, for some examples, the activation map highlights only strong parts of the signal and some parts of the signals are favored for identification.   

\begin{figure}[h]
    \centering
    \begin{tabular}{c}
    \includegraphics[width=0.4\textwidth]{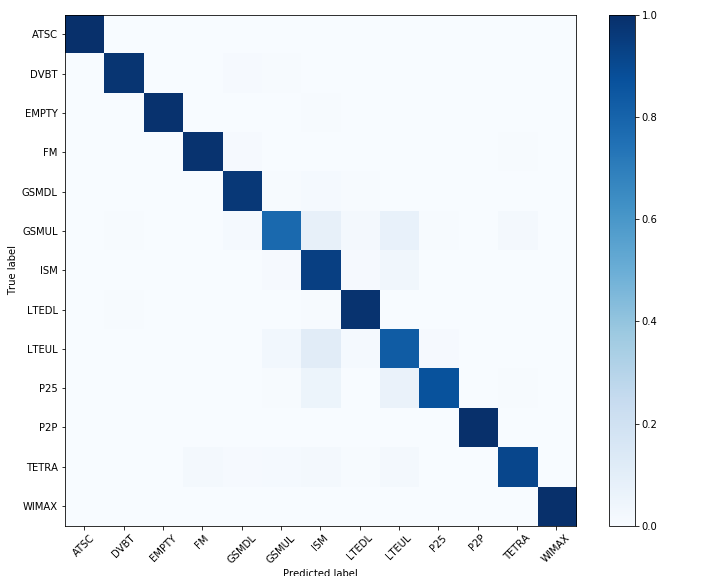}
    \includegraphics[width=0.4\textwidth, trim={0cm 2cm 0cm 0cm}, clip]{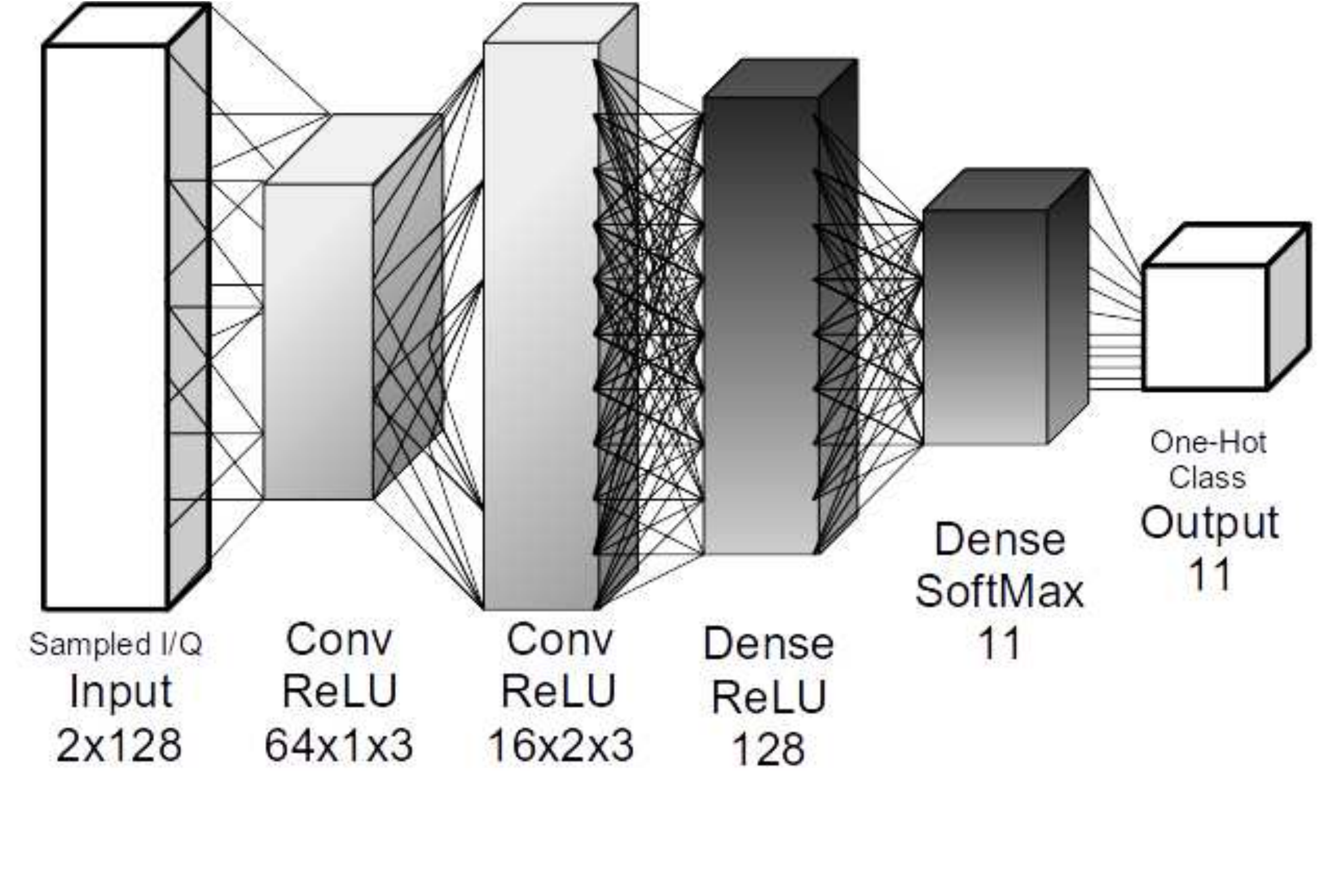}
    \\[-5pt]
    \footnotesize{\hskip-10mm(a)} \hskip67.8mm \footnotesize{(b)} 
    \end{tabular}
\caption{(a) Confusion matrix for RF band classification \cite{OSheaEusipco}, (b) CNN Architecture \cite{RecNetworks}.}
\label{fig:conf_mat_network}
\end{figure}

Fig.~\ref{fig:conf_mat_network} (a) shows the confusion matrix for the classification results. This method for detecting, classifying and localizing emissions within a spectrogram provides reasonable classification performance and reasonable class activation maps corresponding to activity regions in most cases as pictured.  

For the task of supervised modulation recognition, a number of other non-NN based machine learning techniques from literature were compared with that of a convolutional deep learning architecture in terms of performance.  In \cite{RecNetworks}, the generated data set consists of 11 modulations: 8 digital and 3 analog modulations, which are all widely used in wireless communication systems. These consist of BPSK, QPSK, 8PSK, 16-QAM, 64-QAM, BFSK, CPFSK, and PAM4 as digital modulations, and WB-FM, AM-SSB, and AM-DSB as analog modulations.  Data is modulated at a rate of roughly $8$ samples per symbol with a normalized average transmit power of $0$ dB. These signals are exposed to realistic channel effects. Thermal noise results in relatively flat white Gaussian noise at the receiver which forms a noise floor or sensitivity level and SNR. Oscillator drift due to temperature and other semiconductor physics differing at the transmitter and receiver result in symbol timing offset, sample rate offset, carrier frequency offset, and phase difference.  These effects lead to a temporal shifting, scaling, linear mixing/rotating between channels, and spinning of the received signal based on unknown time varying processes. Moreover, real channels undergo random filtering based on the arriving modes of the transmitted signal at the receiver with varying amplitude, phase, Doppler, and delay. This is a phenomenon commonly known as multi-path fading or frequency selective fading, which occurs in any environment where signals may reflect off buildings, vehicles, or any form of reflector in the environment.  

Fig.~\ref{fig:conf_mat_network} (b) shows a simple CNN architecture used for the modulation classification task, an un-tuned 4-layer network utilizing two convolutional layers and two (overly sized) dense fully connected layers.  Layers use ReLU activation functions except for a softmax activation on the output layer to act as a classifier. Dropout regularization is used to prevent over-fitting, while a $ \left \| W \right \|_2 $ norm regularization on weights and $ \left \| \mathbf{h} \right \|_1 $ norm penalty on dense layer activations can also encourage sparsity of solutions \cite{l1l2reg,deconv}.  Training is conducted using a categorical cross-entropy loss and an Adam \cite{AdamOpt} solver.

Expert features (higher order moments, and cumulants) are used by the baseline classifiers. Fig.~\ref{fig:DLvsML} shows the performance results of the Naive Bayes, SVM and CNN network architecture results where the CNN classifier outperforms the Naive Bayes and SVM classifiers at all SNRs. 


For more realistic evaluations, over-the-air dataset was generated in \cite{OSheaJSAC} and the modulation classification performance was compared between \textit{virtual geometry group} (VGG) and \textit{residual networks} (RNs) with better architecture tuning, as well as a stronger XGBoost based baseline. It was shown that the RN approach achieves state-of-the-art  modulation  classification  performance on for both synthetic and  over-the-air signals using datasets consisting of $1$ million examples, each $1024$ samples long.  The RN achieves roughly $5$ dB higher sensitivity for equivalent classification accuracy than the XGBoost baseline at low SNRs while performances are identical at low SNRs. At high SNRs, a maximum classification accuracy rate of $99.8\%$ is achieved by the RN, while the VGG network achieves $98.3\%$ and the baseline method achieves a $94.6\%$ accuracy. 

\subsection{Generative Adversarial Methods for Situation Awareness} \label{sec:GAN}
Radios collect spectrum data samples such as raw (complex-valued) data samples or received signal strength indicator (RSSI) values through spectrum sensing, and use them to train DNNs for various applications such as channel estimation or waveform classification, as discussed in previous sections. There are two important hurdles to overcome before using spectrum data for deep learning purposes. 
\begin{enumerate}
    \item Deep learning requires a large number of data samples to be able to train the complex structures of DNNs. This may not be readily available via spectrum sensing, since a wireless user who spends too much time on spectrum sensing may not have enough time left for other tasks such as transmitting its data packets. Therefore, there may not be enough number of wireless data samples available to train a DNN. \emph{Training data augmentation} is needed to expand the training data collected in spectrum sensing.   
    \item Characteristics of spectrum data change over time as the underlying channels, interference and traffic effects, as well as transmit patterns of wireless users change. Therefore, training data collected for one instant may not be fully applicable in another instant. One example is the channel change when the wireless nodes move from outdoors to indoors, where more multipaths and therefore different channel conditions are expected. \emph{Domain adaptation} is needed to change test or training data collected in spectrum sensing from one domain (e.g., low mobility) to another domain (high mobility).    
\end{enumerate}

\begin{figure}[h]
\centering
\includegraphics[width=0.75\textwidth, trim={3cm 8cm 4cm 4cm}, clip]{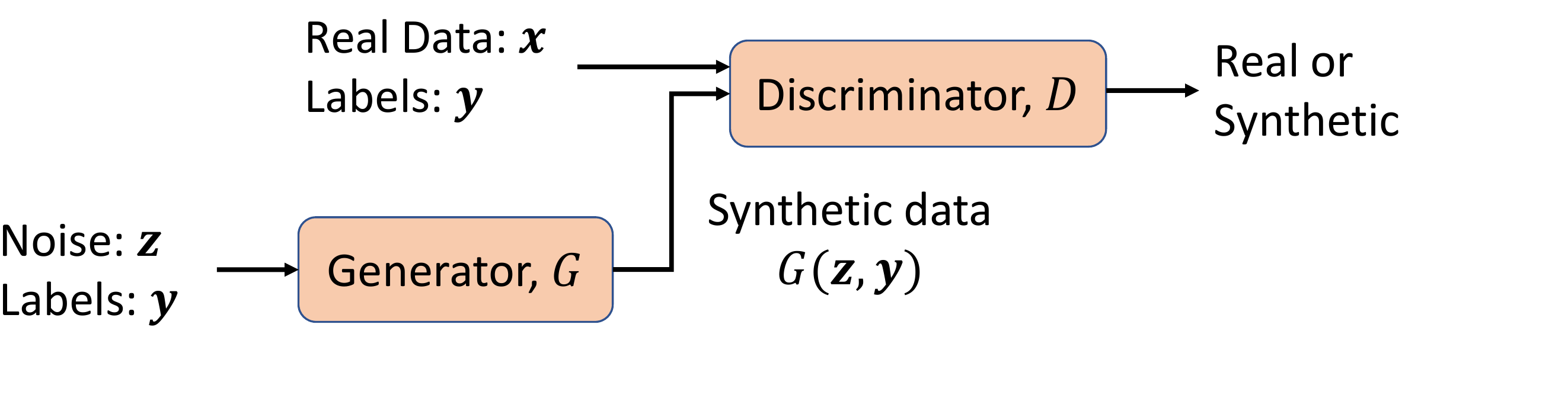}
\caption{Conditional GAN for training data augmentation.}
\label{fig:GANBlockDiagram}
\end{figure}

The GAN has emerged as a viable approach to generate synthetic data samples based on a small number of real data samples in a short learning period and augment the training data with these synthetic data samples for computer vision, text, and cyber applications \cite{Shrivastava2017, Sec4, Sec2}. The GAN consists of a generator and a discriminator playing a minimax game. The generator aims to generate realistic data (with labels), while the discriminator aims to distinguish data generated by the generator as real or synthetic. Conditional GAN extends the GAN concept such that the generator can generate synthetic data samples with labels \cite{Mirza2014}. Fig.~\ref{fig:GANBlockDiagram} shows the conditional GAN architecture. When applied to wireless communications, the GAN needs to capture external effects of channel patterns, interference, and traffic profiles in addition to waveform features. The GAN has been applied for training data augmentation for channel measurements in spectrum sensing \cite{Sec7}, modulation classification \cite{Tang2018}, jamming \cite{Sec1,Sec6}, and call data records for 5G networks \cite{Hughes2019}. 

As an example, consider an adversary that senses the spectrum and observes transmissions of another node (hidden in channel impairments, traffic on/off patterns and other background transmissions). Based on these observations, the adversary trains a DNN to predict when there will be a successful transmission and jams it. See Sec. \ref{sec:security} for details of this setting when deep learning for wireless communications security is discussed. If the adversary waits too long to collect data, it may lose the opportunity to jam transmissions. Therefore, the adversary collects a small number of sensing samples and then augments them through GAN. 

The wireless application of GAN for domain adaptation has remained limited so far. \cite{Sec7} studied the adaptation of training data for spectrum sensing, where a wireless receiver decides if there is an active transmitter (label 1) or not (label 2). There are two environments corresponding to two different channel types, namely Rayleigh fading distributions with variance 0.2 (environment 1) and 2 (environment 2). Assume the receiver has training data for environment 1 and trained a classifier, whereas there is no training data for environment 2. Therefore, the receiver generates synthetic training data samples for environment 2. Training data adaptation consists of a bidirectional GAN \cite{Donahue}, a conditional GAN \cite{Mirza2014}, and a classifier. Bidirectional GAN obtains the inverse mapping from data to the conditioned noise by using a GAN and an autoencoder that together learn to take the inverse of a neural network. As the environment changes from 1 to 2, a new conditional GAN is trained that takes the new samples in environment 2 as real inputs. Instead of random noise as synthetic inputs,  the inverse mapping of the bidirectional GAN is used and the labels in environment 1 is carried to environment 2 to train the CGAN. After CGAN training, a classifier is trained with domain adapted samples and used to label  new samples collected in environment 2. This approach prevents $42\%$ drop in accuracy of SVM-based spectrum sensor operating at $5$ dB SNR \cite{Sec7}.

Separately, the GAN was used in \cite{spoofingGAN} to match waveform, channel, and radio characteristics, and spoof wireless signals that cannot be reliably distinguished from legitimate signals. This attack can be used against signal authentication systems and can be launched to emulate primary user behavior in primary user emulation (PUE) attacks. 

\textbf{Take-away:} This section showed that deep learning provides novel means to characterize and analyze the spectrum. By outperforming conventional machine learning algorithms, DNNs significantly contribute to spectrum situation awareness for channel modeling and estimation with GANs and FNNs and signal detection and classification with CNNs.  

\section{Deep Learning for Wireless Communications Security} \label{sec:security}
Wireless communications are highly susceptible to security threats due to the shared medium of wireless transmissions. A typical example of wireless attacks is the \emph{jamming} attack that aims to disrupt wireless communications by imposing interference at receivers (e.g., see \cite{Jamming}) and causing denial of service (DoS) \cite{DoS}. These attacks use different communication means (e.g., power control \cite{jammingpowercontrol} or random access \cite{jammingrandomaccess}) and  apply at different levels of prior information on attacker's intent \cite{Types}. As radios become smarter by performing more sophisticated tasks, they also become vulnerable to advanced attacks that target their underlying tasks. One example is the \emph{spectrum sensing data falsification} (SSDF) attack, where an adversary that participates in cooperative spectrum sensing deliberately falsifies its spectrum sensing result (namely, whether the channel is busy or idle) \cite{SSDF}. This way, the adversary aims to change the channel occupancy decision from busy to idle (such that the subsequent transmission fails) or from idle to busy (such that no other radio transmits and either the transmission opportunity is wasted or the adversary gets the opportunity to transmit). Data falsification may also occur at other network functions. One example is that routing decisions are manipulated by falsifying measures of traffic congestion (such as queue backlogs) exchanged throughout the wireless network \cite{routingsecurity, INFOCOMsecurity}. 

Beyond these security threats, the increasing use of deep learning by radios  opens up opportunities for an adversary to launch new types of attacks on wireless communications. In particular, deep learning itself becomes the primary target of the adversary. The paradigm of learning in the presence of an adversary is the subject of the emerging field of \emph{adversarial machine learning} \cite{AMLbook} that has been traditionally applied to other data domains such as computer vision. The \emph{exploratory (inference) attack} \cite{Sec9} is one example, where the adversary tries to learn the inner-workings of a machine learning classifier (such as a DNN)  by querying it with some data samples, collecting the returned labels, and building a functionally equivalent classifier. 

Adversarial machine learning provides the necessary optimization mechanisms to launch and mitigate attacks on machine learning. In addition to exploratory attacks, two other popular types of attacks are evasion and causative (poisoning) attacks. In \emph{evasion attacks}, the adversary selects or generates data samples to query a machine learning algorithm such as a deep learning classifier and fool it  into making wrong decisions \cite{Sec8}. In \emph{causative attacks}, the adversary targets the training process and tampers with the training data (i.e., modifies the corresponding labels) such that the machine learning algorithm is not trained adequately \cite{Sec10}. As deep learning is sensitive to errors in training data, this attack is effective against DNNs. While these attacks have been successfully applied in different data domains such as computer vision  (such as image classification \cite{Sec2}) and natural language processing (such as document classification \cite{Sec3}), they cannot be readily applied in wireless communications. The reasons are multi-fold: 
\begin{itemize}
\item The adversary does not have a mechanism to directly query a wireless transmitter but it can only observe its transmission characteristics over the air. 
\item The collection of training data by the adversary is through a noisy channel, i.e., the training data of the adversary is imperfect by default.  
\item The training data and labels of the adversary and its target are different in wireless domain. Their data samples are different because they are received through different channels, whereas their labels are different because their machine learning objectives are different. For example, a transmitter may try to detect whether the channel is busy, while the jammer may try to predict when there will be a successful transmission. 
\end{itemize}

\begin{figure}[!ht]
    \centering
    \includegraphics[width=1\textwidth]{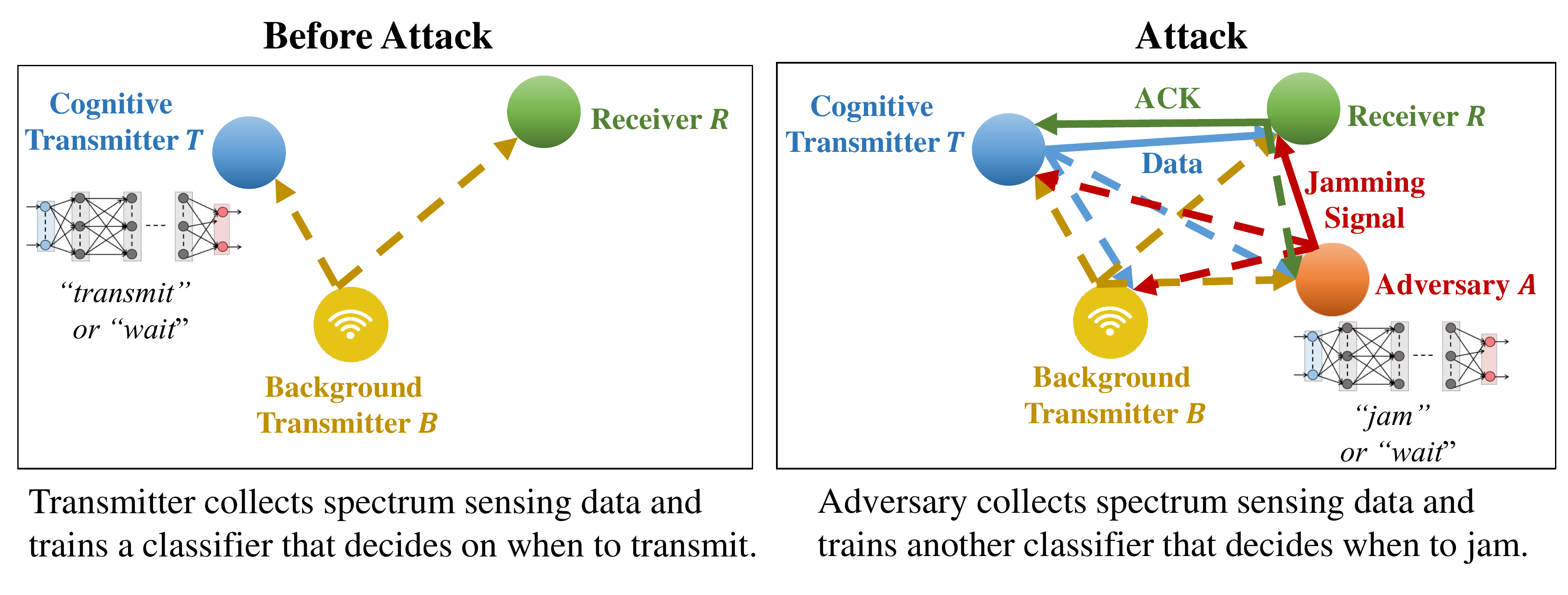}
    \caption{Adversarial deep learning to launch a wireless attack.}
    \label{fig:securityscenario}
\end{figure}

Hence, the application of adversarial machine learning to wireless domain is not trivial and needs to account for the aforementioned differences, both from the attacker and defender perspectives \cite{Sec6, Sec1, Sec5}. As shown in Fig.~\ref{fig:securityscenario}, a basic communication scenario is used to illustrate wireless attacks based on adversarial machine learning \cite{Sec1}. There is one cognitive transmitter $T$ that acts as a secondary user and dynamically accesses the spectrum to communicate with its receiver $R$ while avoiding interference from a background transmitter $B$ that acts as a primary user (e.g., TV broadcast network). $T$ uses a decision function such as a deep learning classifier for its transmissions to capture $B$'s transmission pattern as well as channel effects. There is also an adversary $A$ that does not know the decision function of $T$ 
and tries to learn it by sensing the spectrum. This corresponds to a \emph{black-box} exploratory attack that is followed by other attacks such as jamming to reduce the performance of $T$. In the following, we will describe the exploratory attack on wireless communications and how it is used to launch an effective jamming attack \cite{Sec1}. Then we will present other wireless attacks motivated by adversarial deep learning and discuss defense strategies. 

\subsection{Operational Modes for Transmitter and Adversary}

A synchronized slotted time is assumed where all nodes operate on a single channel (with fixed center frequency and instantaneous bandwidth). Channel gain between any transmitting node $i$ ($T$, $B$, or $A$) and any receiving node $j$ ($R$, $T$, or $A$) is given by $h_{ij}(t)$ in time slot $t$. Then, $j$ receives signal
\begin{eqnarray}
y_j(t) = \sum_{i \in \mathcal{T}(t) }h_{ij}(t) x_i(t) + n_j(t)
\end{eqnarray}
in time slot $t$, where $\mathcal{T}(t)$ is the set of transmitting nodes, $n_j(t)$ is the receiver noise at $j$, and $x_i(t)$ carries a signal if $i \in \mathcal{T}(t)$, otherwise $x_i(t) = 0$. Since channel and noise realizations at $A$ (namely, $h_{BA}(t)$ and $n_A(t)$) and $T$ (namely, $h_{BA}(t)$ and $n_A(t)$) are different, they observe different data input for their tasks. It is assumed that $n_j(t)$ is random according to a zero-mean Gaussian distribution with power normalized as one, and $h_{ij}(t)$ depends on the distance $d_{ij}$ between $i$ and $j$ and type of fading. It is also assumed that signal strength diminishes proportionally to $1/d_{ij}^2$ and log-normal shadowing is used as the shadowing model (namely, flat fading is considered such that the coherence bandwidth of the channel is larger than the bandwidth of the signal and all frequency components of the signal experience the same magnitude of fading). Note that $y_j(t)$ is the signal received during data transmission or sensing periods. In the latter case, $y_j(t)$ is denoted as $s_j(t)$.
Next, the operation modes of background transmitter $B$, transmitter $T$, receiver $R$, and adversary $A$ are discussed, as illustrated in Fig.~\ref{fig:securityscenario}.

\subsubsection{Background transmitter $B$} 
The transmit behavior (idle or busy) of $B$ determines the channel status (idle or busy) in each time slot. There are random packet arrivals at $B$ according to the Bernoulli process with rate $\lambda$ (packet/slot). If $B$ is in idle status and has a packet to transmit, it is activated with certain probability and keeps transmitting until there is no packet anymore in its queue. Since $B$'s busy/idle states are correlated over time, both $T$ and $J$ need to observe not only the last channel status but the past channel states over several time slots to predict the current channel status.

\subsubsection{Transmitter $T$}
In each time slot, $T$ senses the channel and detects whether the channel status is idle or busy, i.e., whether $B$ remains idle or transmits. If idle, $T$ transmits data (a packet) to $R$ in this time slot. $T$ has trained a DNN (unknown to $J$) as the classifier $C_T$ that  classifies the current time slot $t$ as idle or busy based on recent $K_T$ sensing results $(s_T(t-K_T+1), \cdots, s_T(t-1), s_T(t))$. In time slot $t$, the data sample for $C_T$ is
  \begin{equation} \label{eq:featureT}
  \bm{s}_T(t) = (s_T(t-K_T+1), \cdots, s_T(t-1), s_T(t))
  \end{equation}
  and the corresponding label is 
  \begin{equation} \label{eq:labelT}
   L_T(t) = \{``\text{idle}", ``\text{busy}"\},
   \end{equation}
   where ``\text{idle}'' or ``\text{busy}'' means that the channel is idle or busy, respectively.
    Thus, the training data for $C_T$ is built as $\{(\bm{s}_T(t),L_T(t))\}_t$. $T$ obtains the label $L_T(t)$ of a sample only indirectly by observing whether its transmission (if any) is successful or not. A successful transmission indicates an idle channel and a failure indicates a busy channel. Note that this is a noisy observation since a transmission of $T$ may fail or succeed depending on channel conditions even when $B$ does not transmit or transmits, respectively. $T$ deems a transmission as successful if it receives an acknowledgment (ACK) from $R$. If there is no ACK received, then $T$ deems the transmission as failed.  Note that $T$ uses multiple sensing results as its features since features should be able to capture time correlation and help achieve a high sensing accuracy in a short period of time.
      Then classifier $C_T: \bm{s}_T(t) \mapsto L_T(t)$ defines the mapping from  sensing results to occupancy decision and consequently to transmission decision in time slot $t$. 

\subsubsection{Adversary $A$} 
Due to the open nature of wireless spectrum, $A$ can also sense the spectrum and then predict whether there will be a successful transmission (with feedback ACK), or not (without a feedback) in a time slot. In the former case, $A$ transmits to jam the channel in this time slot. In the latter case, $A$ remains idle. Without knowing $C_T$, $A$ builds another classifier $C_A$ itself, which predicts whether there will be a successful transmission, or not, in time slot $t$ based on recent $K_A$ sensing results $(s_A(t-K_A+1), \cdots, s_A(t-1), s_A(t))$. The goal of $A$ is to infer $\mathcal{C}_T$ by building a surrogate classifier $\mathcal{C}_A$. Note that $A$ needs to learn relative channel effects and  $T$'s transmit behavior that in turn depends on $B$'s transmit behavior and corresponding channel effects. This is a difficult learning task that needs to be handled in a black-box manner without any prior knowledge. Therefore, it is imperative for $A$ to use a DNN as $\mathcal{C}_A$.  In time slot $t$, the data sample for $C_A$ is
  \begin{equation} \label{eq:featureJ}
  \bm{s}_A(t) = (s_A(t-K_A+1), \cdots, s_A(t-1), s_A(t))
  \end{equation}
   and the corresponding label is 
   \begin{equation} \label{eq:labelJ}
   L_A(t) = \{``\text{ACK}", ``\text{no ACK}"\},
   \end{equation}
   where ``\text{ACK}" or ``\text{no ACK}" means that there is an ACK following a transmission, or not respectively. Thus, the training data for $C_A$ is built as $\{(\bm{s}_A(t),L_A(t))\}_t$. $C_A$ is defined as the mapping from  sensing results to  prediction of successful transmission and consequently to jamming decision in each time slot. $A$ does not jam all time slots, although doing so can maximize the success of jamming, since $A$ will be easily detected if it is jamming in all time slots due to the high false alarm rate and $J$ may have power budget in terms of the average jamming power (thus it cannot jam all time slots). 

\subsubsection{Receiver $R$}
$R$ receives a transmission of $T$ successfully if the signal-to-interference-and-noise-ratio (SINR) is larger than some threshold $\beta$. SINR captures transmit power, channel, and interference effects. Whenever a transmission is successfully received, $R$ sends an ACK back to $T$ over the short ending period of the time slot. In the meantime, $A$ senses the spectrum and potentially detects the presence of $ACK$ (without decoding it) by considering the fact that ACK messages are typically distinct from data messages (they are short and they follow the data transmission with some fixed time lag).

\subsection{Jamming based on Exploratory Attack}\label{subsec:expatt}

\subsubsection{Deep Learning by Transmitter $T$}
$1000$ samples are collected by $T$ and split by half to build its training and test data. 10 most recent sensing results are used to build one data sample (i.e., $K_T = 10$). $T$ trains an FNN as $C_T$. The Microsoft Cognitive Toolkit (CNTK) \cite{CNTK} is used to train the FNN. $T$ optimizes the hyperparameters of the DNN to minimize $e_T = \max\{ e_T^{MD}, e_T^{FA}\}$, where $e_T^{MD}$ is the error probability for misdetection (a time slot is idle, but $T$ predicts it as busy) and $e_T^{FA}$ is the error probability for false alarm (a time slot is busy, but $T$ predicts it as idle). When the arrival rate $\lambda$ for $B$ is $0.2$ (packet/slot), the optimized hyperparameters of $C_T$ are found as follows. The neural network consists of one hidden layer with $100$ neurons. The cross-entropy loss function  is minimized to train the neural network with backpropagation algorithm. The output layer uses softmax activation. The hidden layers are activated using the sigmoid function. All weights and biases are initialized to random values in $[-1.0,1.0]$. The input values are unit normalized in the first training pass. The minibatch size is $25$. The momentum coefficient to update the gradient is $0.9$.  The number of epochs per time slot is $10$. 

In test time, $\mathcal{C}_T$ is run over $500$ time slots to evaluate its performance. The positions of the $T$, $R$ and $B$ are fixed at locations $(0,0), (10,0)$, and $(0,10)$, respectively. All transmit powers are set $30$ dB above noise power. The SINR threshold $\beta$ is set as $3$. For these scenario parameters, $e_T^{MD}=e_T^{FA}=0$. $T$ makes $400$ transmissions and $383$ of them are successful. Note that $17$ transmissions on idle channels fail due to random channel conditions. Thus, the throughput is $383/500=0.766$ packet/slot and the success ratio is $383/400=95.75\%$. Next, we will show how adversarial deep learning-based jammer can significantly reduce this performance.

\subsubsection{Adversarial Deep Learning by Adversary $A$} 
\emph{Exploratory attack} aims to infer a machine learning (including deep learning) classifier and has been applied to other data domains such as text classification in \cite{Sec9} and to image classification in \cite{Sec8}. In these previous works, the adversary queries the target classifier, obtains labels of a number of samples and then trains a functionally equivalent classifier using deep learning. Two classifiers are functionally equivalent if they provide the same labels for the same sample. However, this approach cannot be applied to the wireless setting due to the differences in data samples and labels. 
\begin{itemize}
\item Data samples at a given time are different, as $T$ and $A$ receive signals through different channels (i.e., due to different distances from $B$ and realizations), such that spectrum sensing results $s_T(t)$ and $s_A(t)$ are different at any time $t$. At a given time $t$, the signal from $B$ is received at $T$, $R$, and $A$ as $y_T(t)=h_{BT}x_B(t)+n_T(t)$, $y_R(t)=h_{BR}x_B(t)+n_R(t)$, and $y_A(t)=h_{BA}x_B(t)+n_A(t)$, respectively where $h_{BT}$, $h_{BR}$, and $h_{BA}$ are the channel gains and $n_T(t)$, $n_R(t)$, and $n_A(t)$ are the receiver noises.  
\item Classifiers of $T$ and $A$ have different types of labels. $T$'s labels indicate whether the channel is busy or idle, whereas $A$'s labels indicate whether $T$ will have a successful transmission, or not. 
\end{itemize}

$A$ trains an FNN as the deep learning classifier $C_A$. For that purpose, $1000$ samples are collected by $A$ and split by half to build its training and test data. $J$ uses the most recent $10$ sensing results to build one data sample (i.e., $K_A = 10$). $J$ aims to jam successful transmissions (with received ACK feedback) only. $A$ optimizes the hyperparameters to minimize $e_A = \max\{ e_A^{MD}, e_A^{FA}\}$, where $e_A^{MD}$ is the error probability for misdetection ($T$'s transmission is successful, but $A$ predicts there will not be an ACK) and $e_A^{FA}$ is the error probability for false alarm ($T$ does not transmit or $T$'s transmission fails (even without jamming), but $A$ predicts that there will be an ACK). The training time (including hyperparameter optimization) is $67$ seconds and the test time per sample is $0.024$ milliseconds. The optimized hyperparameters of the $C_A$ are found as follows. The neural network consists of two hidden layers with $50$ neurons. The cross-entropy loss function is used to train the DNN with backpropagation algorithm. The output layer uses softmax activation. The hidden layers are activated using the hyperbolic tangent (Tanh) function.  All weights and biases are initialized to random values in $[-1.0,1.0]$. The input values are unit normalized in the first training pass.  The minibatch size is  $25$. The momentum coefficient to update the gradient is $0.9$.  The number of epochs per time slot is $10$. With these hyperparemeters, the error $e_A$ is minimized to $1.48\%$. Note that the hyperparameter optimization affects the accuracy. For instance, if the number of layers is decreased to $1$, the error $e_A$ increases to $1.73\%$. Similarly, if the number of neurons per layer is changed to $30$, the error $e_A$ increases to $2.22\%$.

In test time, $\mathcal{C}_A$ is run over $500$ time slots to evaluate its performance. The position of $A$ is fixed at location $(10,10)$ and its jamming power is $30$ dB above noise power. If there is no jamming, $T$ will have $383$ successful transmissions. Under $A$'s attack, the number of misdetections is $16$, i.e., misdetection probability is $e_A^{MD}=16/383=4.18\%$ (majority of successful transmissions are jammed), and the number of false alarms is $17$, i.e., false alarm probability is $e_A^{FA}=17/(500-383)=14.53\%$.
As the significant impact of this attack, there are only $25$ successful transmissions among $400$ transmissions. Thus, the throughput of $T$ is reduced from $0.766$ packet/slot to $25/500=0.05$ packet/slot and the success ratio of $T$ is reduced from $95.75\%$ to $25/400=6.25\%$.

As a benchmark, a conventional attack without adversarial deep learning is also considered. In this \emph{sensing-based jamming}, $A$ jams the channel if its received power  during spectrum sensing in the current slot is greater than a threshold $\tau$. Note that the performance of a sensing-based jammer relies on proper selection of $\tau$. If $\tau$ is too low, the number of false alarms increases. If $\tau$ is too high, then the number of misdetections increases. Note that $\tau$ is usually given as a fixed value since there is no clear mechanism to select $\tau$. For a performance upper bound, $\tau$ is selected as $3.4$ that minimizes $e_A$ and used to compute the throughput and the success ratio of the transmitter in the presence of sensing-based jammer. Then $e_A^{MD}=12.8\%$ and $e_A^{FA}=12.6\%$. Note that $e_A^{MD}$ grows quickly to $30.0\%$ when $\tau$ is increased to $5$, whereas $e_A^{FA}$ grows to $14.0\%$ when $\tau$ is reduced to $2$. With the best selection of $\tau$, the throughput of $T$ is reduced to $0.140$ packet/slot and the success ratio of $T$ is reduced from $16.99\%$. On the other hand, if $\tau$ is selected arbitrarily (say, $4.7$), the throughput of $T$ becomes $0.576$ packet/slot and the success ratio of $T$ becomes $69.90\%$ (i.e., the attack is not as effective). The results which are summarized in Table~\ref{table:attacksummary} show the importance of adversarial deep learning in launching wireless jamming attacks. 

\begin{table}
\caption{Effect of different attack types on the transmitter's performance \cite{Sec1}.}
\centering
{\small
\begin{tabular}{c|c|c}
Attack type & Throughput & Success ratio \\ \hline \hline
No attack & $0.766$ & $95.75\%$ \\ \hline
Adversarial deep learning & $0.050$ & $6.25\%$ \\ \hline
Sensing-based attack ($\tau = 3.4$) & $0.140$ & $16.99\%$ \\ \hline
Sensing-based attack ($\tau = 4.7$)  & $0.576$ & $69.90\%$
\end{tabular}
}
\label{table:attacksummary}
\end{table}

\subsubsection{Generative Adversarial Learning for Wireless Attacks} \label{sec:GAN2}
In the training process of adversarial deep learning, $A$ collected $500$ samples to build its classifier $C_A$. From a practical attack point of view, it is critical to shorten this initial learning period of $A$ before jamming starts. For that purpose, $J$ builds the GAN to generate synthetic data samples based on a small number of real data samples in a short learning period. Then it uses these synthetic data samples to augment its training data, as discussed in Sec. \ref{sec:GAN}. 

\begin{figure}
  \centering
  \includegraphics[width=0.5\columnwidth]{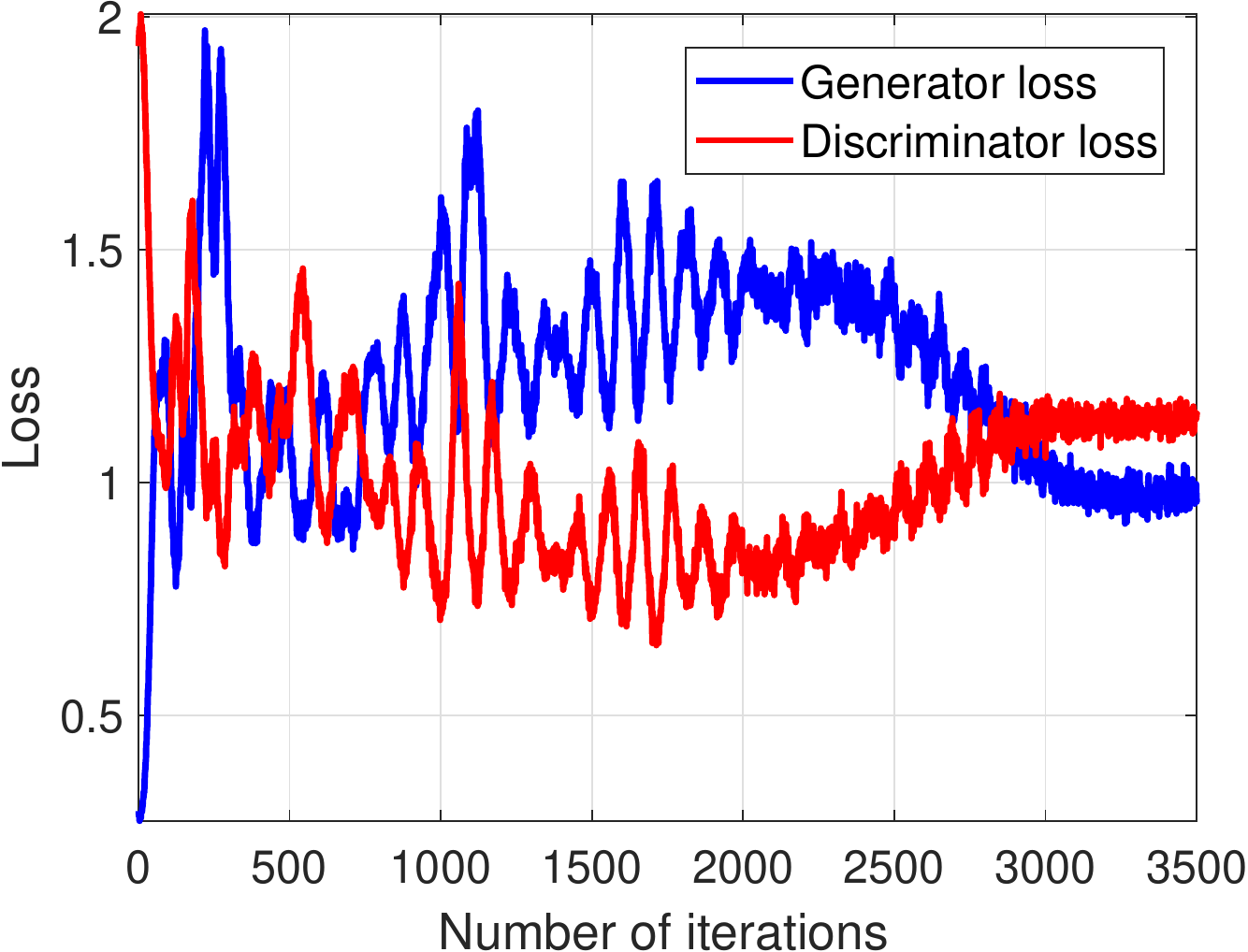}
  \caption{Discriminator and generator losses during training \cite{Sec1}.}\label{fig:dgloss}
\end{figure}

The conditional GAN is implemented in TensorFlow \cite{Tensorflow} by using the FNNs with three hidden layers each with $128$ neurons for both generator and discriminator of the GAN. Leaky ReLu is used as the activation function. Adam optimizer \cite{AdamOpt} is used as the optimizer to update the weights and biases. The output of each hidden layer is normalized (via batch normalization). Fig.~\ref{fig:dgloss} shows the losses of generator and discriminator. Note that the losses fluctuate significantly when the GAN training starts and eventually converges after 3000 iterations of the GAN training process.  

The similarity between real and synthetic data distributions are measured by the Kullback-Leibler (KL) divergence. The KL divergence is given by 
\begin{eqnarray}
D_{\text{KL}} (P \| Q) = - \sum_{x \in \mathcal{X}} P(x) \log  \left( \frac{Q(x)}{P(x)} \right)
\end{eqnarray}
for two distributions $P$ and $Q$ with the support over $\mathcal{X}$. Denote $P$ as the distribution of synthetic data samples (generated by the GAN), $Q$ as the distribution of real samples, and $c$ as the random variable for the channel status ($c = 0$ if idle and $c = 1$ if busy). Define $P_i(x) = P(x | c = i)$ for $i=0,1$. Then, $D_{\text{KL}} (P_0 \| Q_0) = 0.1117$ and $D_{\text{KL}} (P_1 \| Q_1) = 0.1109$. The test time per sample is measured as $0.024$ milliseconds (much smaller than the channel coherence time). If sensing results are obtained per second and $500$ measurements are made, it takes $500$ seconds to collect $500$ RSSI levels without using the GAN. It takes $23$ seconds to train the GAN  using a GeForce GTX 1080 GPU and generate $500$ synthetic samples from the GAN.  Since $10$ real samples are collected over $10$ seconds, it takes $33$ seconds to prepare data with the GAN. Hence, the GAN significantly reduces the data collection time before $A$ starts jamming.
When $A$ builds its classifier $C_A$ based on $10$ real data samples, the error probabilities are $19.80\%$ for false alarm and $21.41\%$ for misdetection. After adding $500$ synthetic data samples, the error probabilities drop to $7.62\%$ for false alarm and to $10.71\%$ for misdetection, namely close to the levels when $500$ real data samples are used to train the DNN.

\subsection{Other Attacks based on Adversarial Deep Learning}
\begin{figure}[!ht]
    \centering
    \includegraphics[width=0.8\textwidth]{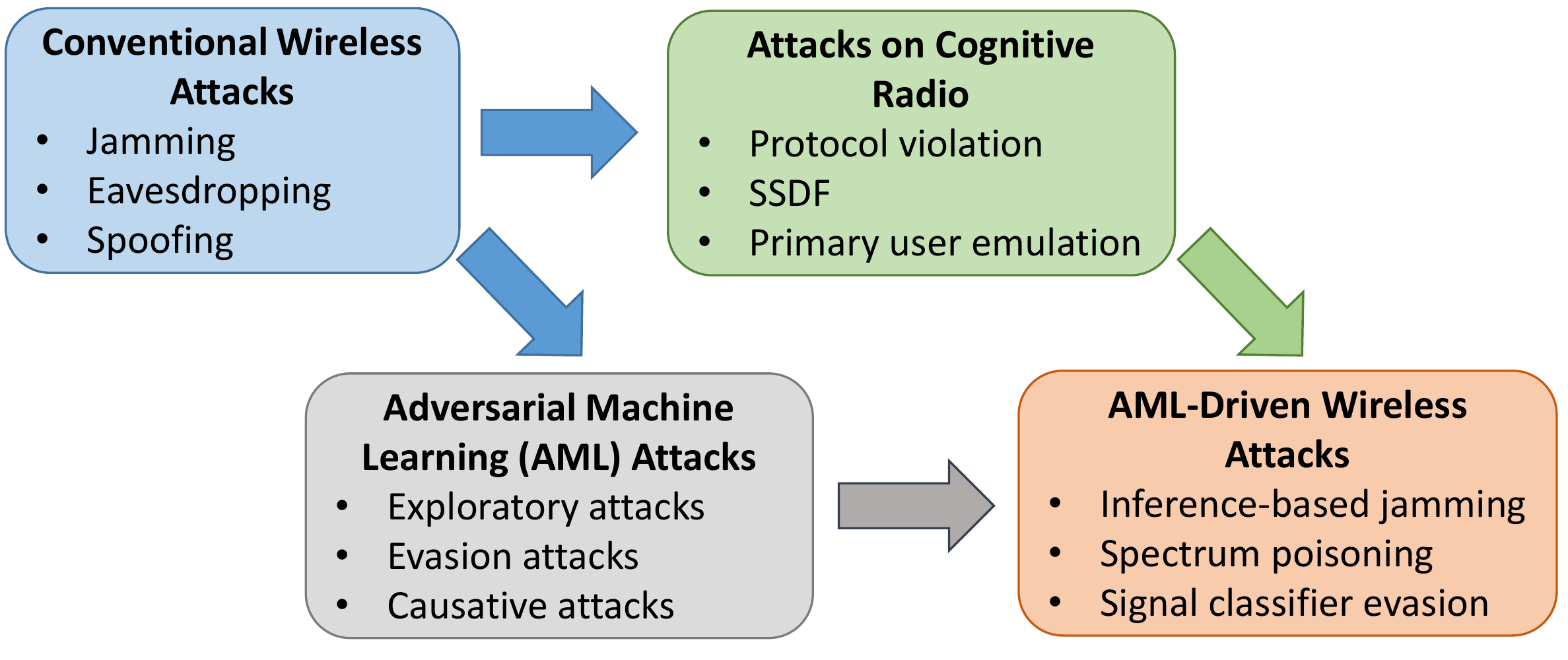}
    \caption{From conventional wireless attacks to adversarial machine learning.}
    \label{fig:security_taxonomy}
\end{figure}

There are various other wireless attacks that can be launched through adversarial machine learning. A brief taxonomy of attacks from the conventional settings to adversarial machine learning is shown in Fig.~\ref{fig:security_taxonomy}. 

\subsubsection{Spectrum Poisoning Attack}
Adversarial deep learning can be also used to launch \emph{over-the-air spectrum poisoning attacks} \cite{Sec5}. Using the results of exploratory attack, the adversary falsifies the transmitter's spectrum sensing data over the air by transmitting during transmitter's short spectrum sensing period. Depending on whether the transmitter uses the sensing data as test data to make transmit decisions or for retraining purposes, either it is fooled into making incorrect decisions (evasion attack), or the transmitter's algorithm is retrained incorrectly (causative attack). Both attacks substantially reduce the transmitter's throughput. Note that these attacks differ from the SSDF attack, since the adversary does not participate in cooperative spectrum sensing and does not try to change channel status labels  directly.  Instead,  the  adversary  injects  \emph{adversarial  perturbations}  to the  channel  and aims to fool  the  transmitter  into  making  wrong spectrum access  decisions. A defense scheme can be applied by the transmitter that deliberately makes a small number of incorrect transmissions (selected by the confidence score on channel classification) to manipulate the adversary's training data. This defense effectively fools the adversary and helps the transmitter sustain its throughput \cite{Sec5}.

Another attack that targets spectrum sensing is \emph{priority violation} attack \cite{IoTAML}, where the adversary transmits during the sensing phase by pretending to have higher priority (e.g., emulating primary user behavior) and forces a target transmitter into making wrong decisions in an evasion attack.

\subsubsection{Evasion Attack Against Signal Classifiers}
\emph{Adversarial perturbations} can be added to data samples in the test phase for other wireless communications tasks such as signal classification \cite{LarssonAML, Headley19, Deniz19, Silvija19}. In this \emph{evasion attack}, a receiver aims to classify the incoming signals with respect to waveform characteristics. In the meantime, an adversary transmits as well such that a carefully controlled interference signal is added to the received signal and causes the classifier to misclassify the received signal. This problem was studied in \cite{LarssonAML, Headley19} for modulation classification using a CNN-based classifier. Both white-box and black-box attacks on the deep learning classifier are shown to be effective in terms of increasing the classification error with small over-the-air perturbations added to the received signal. \cite{Deniz19, Silvija19} developed means to prevent an intruder from successfully identifying the modulation scheme being used.

Overall, the attacks that target spectrum sensing or signal classification transmit short signals with low power. Therefore, they are  more energy efficient and harder to detect compared to conventional attacks that jam the long data transmission period.

\subsubsection{Deep Learning-based Defense Against Wireless Threats}
In addition to adversarial deep learning, wireless security threats have been studied with defense mechanisms based on deep learning. Against jamming attacks, \cite{TwoDim} developed a deep Q-network algorithm for cognitive radios to decide whether to leave an area of heavy jamming or choose a frequency-hopping pattern to defeat smart jammers. \cite{Wu2017} trained a CNN network to classify signals to audio jamming, narrowband jamming, pulse jamming, sweep jamming, and spread spectrum jamming. \cite{gunes} applied a wavelet-based pre-processing step that highlights the disrupted parts of the signal before classifying signals as jammers using a CNN. 
Another example is \emph{signal authentication} with deep learning as an IoT application. \cite{Saad2018} presented a deep learning solution based on  a long short-term memory (LSTM) structure to extract a set of
stochastic features from signals generated by IoT devices and dynamically watermark these features into the signal. This method was shown to effectively authenticate the reliability of the signals.

\subsection{Defense Against Adversarial Deep Learning}
A typical first step of adversarial deep learning is the exploratory attack where $A$ builds the surrogate classifier $\mathcal{C}_A$ to infer the transmit behavior of $T$. An effective defense follows from disrupting the training process of $\mathcal{C}_A$. In this defense, $T$ does not always follow the labels returned by $\mathcal{C}_A$ and changes them for some of its data samples when making transmit decisions \cite{Sec1}. In particular, $T$ changes the label  ``ACK'' (i.e., ``a successful transmission'') to ``No ACK'' (i.e., ``no successful transmission''), and vice versa. This way, $A$'s training data is manipulated and $A$ cannot build a reliable classifier in the exploratory attack. As $T$ poisons the training process of $A$ by providing wrong training data, this defense corresponds to a \emph{causative} (or \emph{poisoning}) attack of $T$ back at $J$. By deliberately taking wrong decisions in certain time slots, $T$ does not transmit even if channel is predicted as idle, and transmits even if channel is predicted as busy.

While this defense increases the uncertainty at $A$, there is a trade-off in the sense that wrong transmit decisions would reduce the transmission success of $T$. Therefore, $T$ needs to decide to flip its decision in a small number of carefully selected time slots. Let $p_d$ denote the percentage (\%) of time slots in which $T$ decides to flip labels. $p_d$ is considered as a defense budget. $T$ uses the likelihood score $S_T(t)$ (namely the likelihood of whether a channel is idle) returned by DNN to decide when to take the defense action.
If $S_T(t)$  is less than a threshold $\eta$, $T$ classifies a given time slot $t$ as idle; otherwise $T$ classifies it as busy. When $S_T(t)$ is far away from $\eta$, then such a classification has a high confidence; otherwise the confidence is low. For the FNN structure used in previous subsection, $\eta =  0.25$, which is chosen to minimize $e_T$. 
To optimize the defense mechanism, $T$ performs defense operations in a time slot $t$ when $S_T(t)$ is  close to $0$ or $1$, since $T$'s transmission decisions are more predictable in such a time slot. Subject to $p_d$ values, $T$ changes labels in different time slots and $A$ ends up building different classifiers with different hyperparameters compared to the previous case of no defense.

\begin{figure}[h]
    \centering
    \begin{tabular}{c}
    \includegraphics[width=0.5\textwidth,trim={0cm 0cm 0cm 0.5cm},clip]{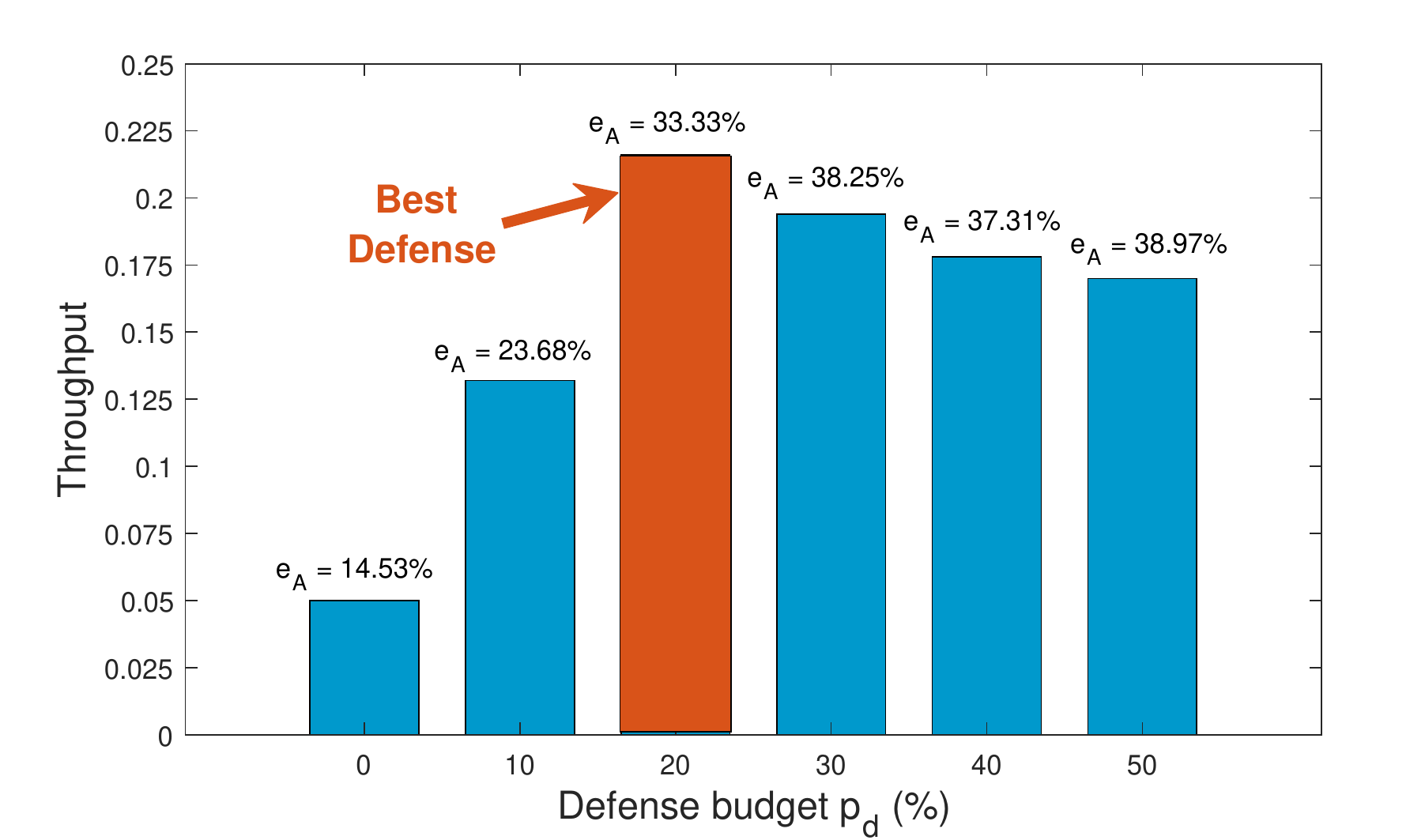}
    \includegraphics[width=0.5\textwidth,trim={0cm 0cm 0cm 0.5cm},clip]{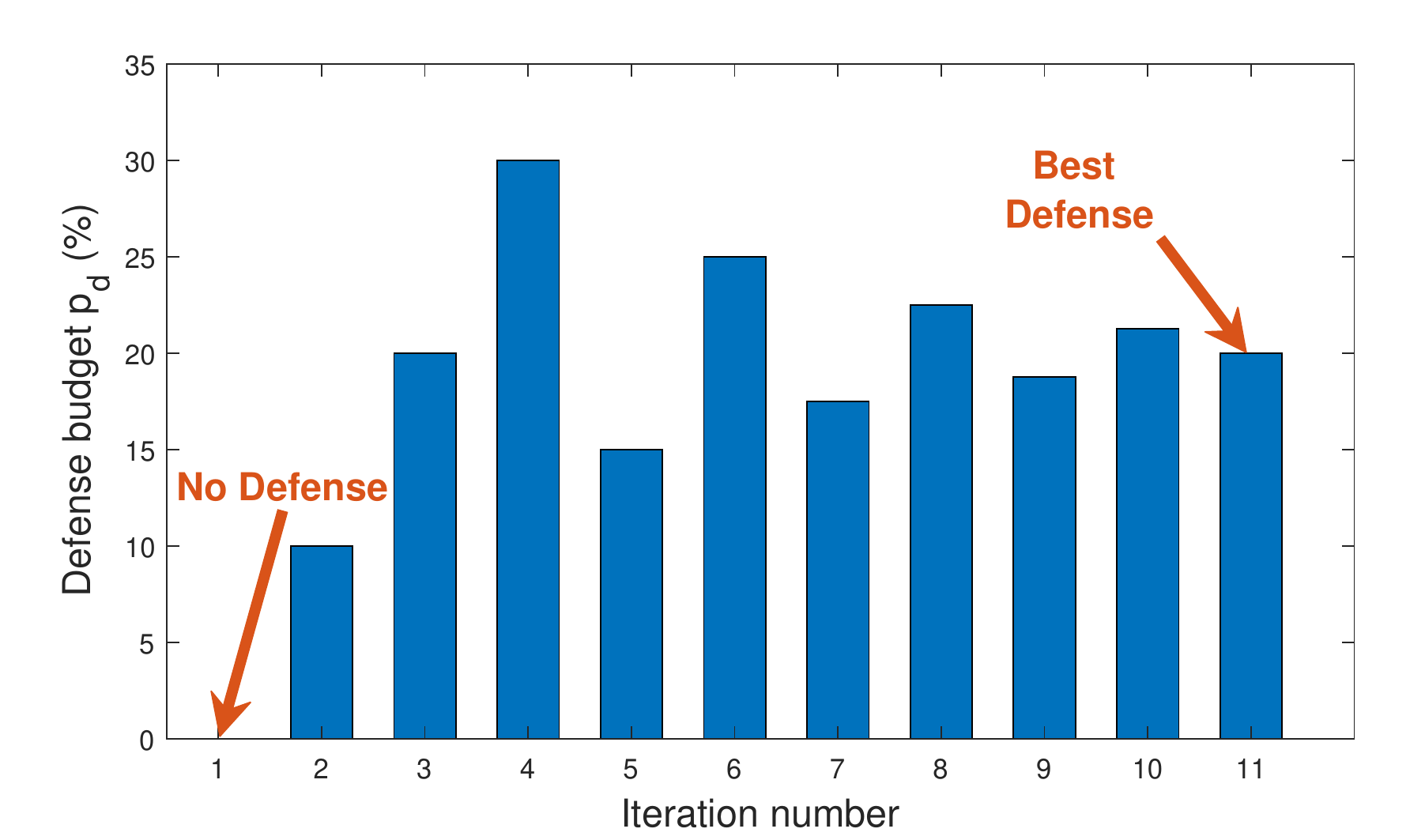}
    \\[-5pt]
    \footnotesize{\hskip9mm(a)} \hskip86.8mm \footnotesize{(b)} 
	\end{tabular} 
\caption{ (a) Effects of transmitter's defense on the adversary \cite{Sec1}, (b) Dynamic adaptation of transmitter's defense against the adversary \cite{Sec1}.}
\label{fig:defense1_2}
\end{figure}

Fig.~\ref{fig:defense1_2} (a) shows the results when $T$ operates with different defense budgets. As $p_d$ increases, $A$'s error probabilities and $T$'s throughput start increasing significantly. $T$'s throughput reaches maximum when $p_d = 10\%$. As $p_d$ increases further, the growth in $A$'s error probabilities saturates and cannot compensate the errors in channel access decisions anymore. As a result, $T's$ throughput starts decreasing. To determine the best value of $p_d$, $T$ can start attack mitigation with a fixed level of $p_d$ and then gradually increase or decrease $p_d$ in response to changes in its throughput that is measured through the received ACK messages. Fig.~\ref{fig:defense1_2} (b) shows how $p_d$ is adapted over time to optimize the throughput. 

\textbf{Take-away:} This section showed that deep learning can be effectively used in an adversarial setting to launch successful attacks to reduce communication performance. In turn, the adversary can be fooled by manipulating its sensing data samples at certain time instances that are selected by deep learning prediction results. 

\section{Conclusion} \label{sec:conclusion}
Deep learning has made rapid strides in addressing unique challenges encountered in wireless communications that need to learn from and adapt to spectrum dynamics quickly, reliably, and securely. We presented the recent progress made in applying deep learning to  \emph{end-to-end (physical layer) communications}, \emph{spectrum situation awareness}, and \emph{wireless security}.  First, we discussed how to formulate transmitter and receiver design at the physical layer as an autoencoder that is constructed as DNNs. We showed that this formulation captures channel impairments effectively and improves performance of single and multiple antenna, and multiuser systems significantly compared to conventional communication systems. Second, we showed that deep learning can help with channel modeling and estimation as well as signal detection and classification when model-based methods fail. The GAN can be applied to reliably capture the complex channel characteristics for the purpose of channel estimation or spectrum data augmentation, while CNNs can improve the signal classification accuracy significantly compared to conventional machine learning techniques. Third, we discussed the application of adversarial deep learning to launch jamming attacks against wireless communications. Starting with an exploratory attack, the adversary can use DNNs to reliably learn the transmit behavior of a target communication system and effectively jam it, whereas a defense mechanism can fool the adversary by poisoning its DNN training process. 

The research topics discussed in this chapter illustrated key areas where deep learning can address model and algorithm deficits, enhancing wireless communications. The progress so far clearly demonstrated that deep learning offers new design options for wireless communications and enhances spectrum situational awareness, while adversarial use of deep learning poses an emerging threat to wireless communications and casts communications and sensing into an interesting adversarial game. Numerous additional deep learning applications in wireless communications are on the horizon, which will potentially change the way we model, design, implement, and operate new generations of wireless systems, and shift the field to be more data-centric than ever before.

\end{document}